\documentclass{aastex631}

\usepackage{booktabs, comment}
\usepackage{threeparttable}
\begin{document}

\title{Numerical \texttt{AXP4} Simulations of Pulse Profiles for Binary Accreting X-ray Pulsars -- II: \\A Case Study of Centaurus X-3}

\correspondingauthor{Parisee S. Shirke}
\email{parisee@iucaa.in}

\author[0000-0003-2977-3042]{Parisee S. Shirke}
\affiliation{Inter-University Centre for Astronomy and Astrophysics \\
Post Bag 4, Ganeshkhind, \\
Pune 411 007, India}

\correspondingauthor{Gulab C. Dewangan}
\email{gulabd@iucaa.in}

\author[0000-0003-1589-2075]{Gulab C. Dewangan}
\affiliation{Inter-University Centre for Astronomy and Astrophysics \\
Post Bag 4, Ganeshkhind, \\
Pune 411 007, India}

\begin{abstract}
The pulse shapes simulated in the accompanying paper Part -- I are compared with observations of a model binary accreting X-ray pulsar, Centaurus X-3. With known Cen X-3 inclination angles provided as input to the \texttt{AXP4} code, the generated pulse profile is suitably compared with the corresponding observed energy-resolved \textit{AstroSat}/LAXPC pulse profile. The pulsed fraction is proposed as a robust, quantitative measure for estimating the size of the emission region of Centaurus X-3 by extending the simulations to include spherical caps of varying fractional surface coverage of the neutron star -- over the full range of $0-100\%$, up to very large caps (with polar half angle $> 30$\textsuperscript{o}). The hotspot radius thus derived drops by an order of magnitude from 12.27 km to $1.36^{+0.29}_{-0.26}$ km, within the ballpark of the standard model value of $\sim$1 km, after including the effect of gravitational light bending, lending further weight to its routine emphasis in the literature. The energy- and luminosity-dependence of the composite gravitationally bent and slab-integrated pulse profiles is further studied. As the pulse profile is sensitive to luminosity variations, the correlation of the size of a finite polar cap and its dependence on X-ray luminosity -- through the rate and subsequently, the geometry of accretion -- is discussed. Although a single, model pulsar was chosen for this work to exhibit the depth of the physical and astrophysical prospects of such a probe, this exercise can be extended to a wide range of existing X-ray pulse profiles of known binary accreting pulsars available in galactic catalogs, especially, with the possible inclusion of accretion columns (with cylindrical co-ordinate transformation) in the future. 
\end{abstract}


\section{Introduction} \label{sec:Intro}

The interplay of various participating physical factors namely, accretion rate, geometry (size and shape) of the emission regions,  X-ray luminosity, observing X-ray energy band, source magnetic field (local strength and structure) $\vec{B}$ configuration governs the final X-ray emission from a binary accreting X-ray pulsar based on the resultant radiation-matter interactions that occur in the plasma atmosphere (See \cite{2022arXiv220414185M} and \cite{tauris2023physics} for a recent subject review). \cite{2021MNRAS.501..109C, 2021MNRAS.501..129C} have enlisted some additional physical effects that can come into play which makes pulsars promising galactic laboratories for studying extreme physics. 
This renders pulse profile modeling a complex exercise. A full, coherent understanding of the underlying physics of X-ray formation is yet to be achieved although thorough discussions can be found in \citeauthor{MeszarosNagel1985a}'s works. The accompanying paper (Part -- I) is rooted in these early and model attempts with an aim to provide new, updated, high-resolution simulations through the development of the \texttt{Accreting X-ray Pulsar Pulse Profile Producer} (\texttt{AXP4}) code in Python \texttt{3} along with the \texttt{GR Light Bender} and \texttt{Composite Slab Integrator} modules. These can handle several source pulsar configurations.

Instead of democratic simulations of a large sample of pulsar pulse profiles resulting in a series of case-to-case comparisons, the choice of a model pulsar helps probe the generalized behaviour of accreting X-ray pulsars to a greater depth. This retains the focus of the exercise on studying the physics of the pulsars themselves rather than demonstrating the efficacy of a pulse profile-generating algorithm (needless to say, any physical insight derived from the development of such algorithms will remain useful).

Centaurus X-3 is a typical, widely studied high mass X-ray binary source (extensive literature spanning the full timeline of X-ray astronomy from 1971 -- 2024 is available) and is chosen here as a model binary accreting X-ray pulsar. Apart from its nearly circular orbit \citep{2009A&A...507.1211R, 2021JApA...42...58S}, it also presents a high X-ray flux with a mixed pencil and fan behaviour depending on its luminosity state. Apart from the distorted magnetic dipole \citep{1996ApJ...467..794K, 2008ApJ...675.1487S}, it does not show major peculiar oddities or prominent deviations from general behaviour.

Centaurus X-3 is known to have its rotation axis inclined at $65$\textsuperscript{o} to the line of sight of an observer on Earth and a magnetic dipole inclined at $20$\textsuperscript{o} with respect to the rotation axis \citep{2000ApJ...530..429B, 1989PASJ...41....1N, 1981A&A...102...97W}. Timing analysis of its 3.0 -- 80.0 keV light curve observed by \textit{AstroSat}/LAXPC from 12 -- 13 December, 2016 was performed in \cite{2021JApA...42...58S}. Cen X-3 is known to exhibit a mixed pencil and fan beam emission across different luminosity states (See e.g. \cite{bachhar2022timing}). While the former points to a slab emission region corresponding to lower accretion rates, the latter emanates from an accretion column overlying the polar caps that supports high mass accretion rates \citep{Schonherr}. In particular, \cite{1996ApJ...467..794K} report two equal and almost antipodal emission regions at the magnetic poles with a hollow cone emission geometry \citep{2001ApJ...563..289K}. Although other high mass X-ray binaries like Cen X-3 are favourable candidates for fan beam emission, occurrences of slab geometries are expected in large number from low mass X-ray binaries. Independent efforts in neutron star research have recently put forth the possibilities of various hotspot configurations (See e.g. \cite{2019ApJ...887L..21R, miller2019psr, miller2021radius} for investigations with \textit{NICER} and \textit{XMM-Newton} missions).

In this paper, numerical simulations of pulse profiles for binary accreting X-ray pulsars from the accompanying paper are compared with existing observations towards the possible geometric characterization of the source. Previous attempts have been made by \cite{1981A&A...102...97W} and \cite{1991MNRAS.251..203L}. Also, see similar works based on \cite{MeszarosNagel1985a}, e.g. \cite{RiffertMeszaros1988, 1993ApJ...406..185R, 1994A&A...286..497P, 1995MNRAS.277.1177L, 2001ApJ...563..289K, 2004ApJ...613..517L, 2007ESASP.622..403T}). 

Various mathematical phenomenological approaches to pulse profile modeling involve (i) fitting with polynomials of trigonometric functions $\cos^n$x and $\sin^n$x, $n=1-3$ \citep{1993ApJ...406..185R, 1994A&A...286..497P} and (ii) Fourier series expansions (See e.g. \cite{1989PASJ...41....1N, giacconi1971discovery}). In this paper, the pulse shape is fitted with (iii) direct ground-up physical modeling of the pulsar rotation building on the work of \cite{MeszarosNagel1985a}. 
The paper particularly focusses on the effects of the much-emphasized general relativistic considerations \citep{2022arXiv220414185M, Meszaros1988}. (Also, see \cite{2018A&A...613A...2B} and references therein for detailed studies in the case of fellow compact stars -- black holes) and that of an increasing slab area -- to geometrically spherical caps on the neutron star. 
One of the primary features to be taken into account for such modeling is the effect of gravitational bending of light under the influence of a strong gravitational field around the compact star \citep{RiffertMeszaros1988,1995MNRAS.277.1177L, 2001ApJ...563..289K, 2004ApJ...613..517L, 2007ESASP.622..403T} as confirmed in Sec. \ref{g}. 

The effect of a finite polar cap size and its dependence on energy and luminosity is another aspect  \citep{1981A&A...102...97W} that is investigated in Secs. \ref{depend} and \ref{energy}. (See \cite{leahy1990x, cappallo2017geometry} for further discussions.)
This is achieved with a new, custom \lq \texttt{Surface Coverage}' module and allows for a variation in the ratio of the emission area to the full neutron star spherical surface. This is swept at a high polar angular resolution of $0.5\%$ and for the sake of theoretical completeness, the full span of surface coverage, 0 -- 100$\%$. For comparison, we supplement these profiles generated using \citeauthor{MeszarosNagel1985a}'s emission beams with alternative pulse profiles generated using hypothetical, fiducial input injection from uniform, isotropic test emitters.

\section{Simulations for Centaurus X-3}

\begin{figure}
    \centering 
    \includegraphics[trim=5 8 5 5, clip=true, height=0.45\textheight]{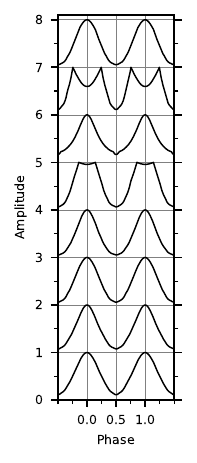}
    \hspace{-.3cm}
    \includegraphics[trim=25 8 5 5, clip=true, height=0.45\textheight]{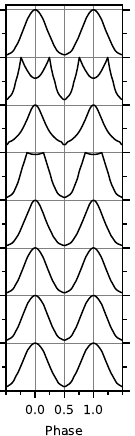}
    \hspace{-.3cm}
    \includegraphics[trim=25 8 5 5, clip=true, height=0.45\textheight]{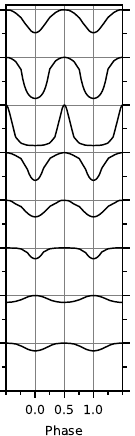}
    \hspace{-.3cm}
    \includegraphics[trim=25 8 5 5, clip=true, height=0.45\textheight]{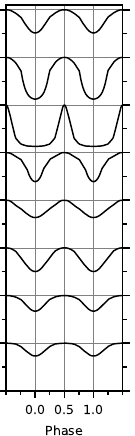}\\ 
    \includegraphics[trim=5 8 5 5, clip=true, height=0.45\textheight]{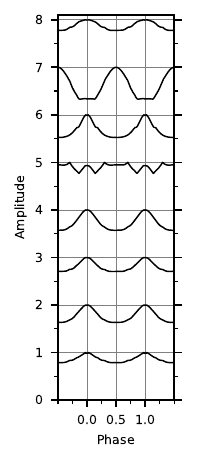}
    \hspace{-.3cm}
    \includegraphics[trim=25 8 5 5, clip=true, height=0.45\textheight]{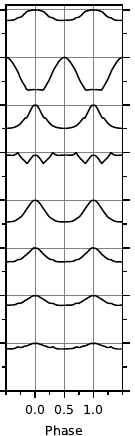}
    \hspace{-.3cm}
    \includegraphics[trim=25 8 5 5, clip=true, height=0.45\textheight]{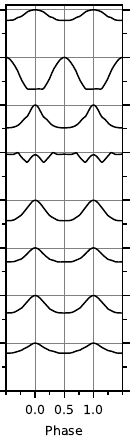}
    \hspace{-.3cm}
    \includegraphics[trim=25 8 5 5, clip=true, height=0.45\textheight]{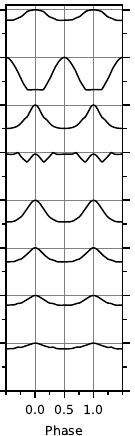}
    \caption{Simulated normalized Centaurus X-3 pulse profiles as in Figs. $4-7$ and 9 of the accompanying paper for the known ($i_1=65\textsuperscript{o},~i_2=20\textsuperscript{o}$) set of inclination (\textit{top row}) for a point hotspot with model beaming functions for (\textit{first column}) shallow slab, (\textit{second column}) deep slab, (\textit{third column}) shallow column (\textit{fourth column}) deep column for the X-ray photon energies \textit{(eight individual panels, bottom to top)} 1.6, 3.8, 9.0, 18.4, 29.1, 38.6, 51.7 and 84.7 keV and (\textit{bottom row}) including the effect of gravitational light bending with the corresponding gravitationally red-shifted X-ray photon energies 1.3, 3.1, 7.5, 15.2, 24.1, 32.0, 42.8 and 70.2 keV (\textit{first column}) for shallow slab and (\textit{second column}) deep slab and the composite effect of light bending and surface integration (\textit{third column}) for shallow slab and (\textit{fourth column}) deep slab at a spin phase resolution of \texttt{0.001}.}
    \label{model_simu}
\end{figure}

\cite{2000ApJ...530..429B} report the inclination angles of the line of sight ($i_1$) and magnetic dipole axis ($i_2$) with respect to the rotation axis as 65$\textsuperscript{o}$ and 20$\textsuperscript{o}$, respectively, which are in agreement with \cite{1989PASJ...41....1N} and \cite{1981A&A...102...97W}, respectively. Note that the values of $i_1$ and $i_2$ are inter-changeable inputs, as the resultant profiles are mathematically identical in either case \citep{MeszarosNagel1985b}. These are fed as inputs to the \texttt{AXP4} simulator developed in Part -- I of this paper. The simulator provides models for 8 discrete values of red-shifted X-ray photon energies 1.3, 3.1, 7.5, 15.2, 24.1, 32.0, 42.8 and 70.2 keV (with the cyclotron line lying at a red-shifted 32 keV for a dipolar magnetic field strength of $10^{12}$ Gauss) as shown in Fig. \ref{model_simu}. 

\begin{figure}
    \centering
    \includegraphics[trim=0 238 970 258, clip=true, width=.32\textwidth]{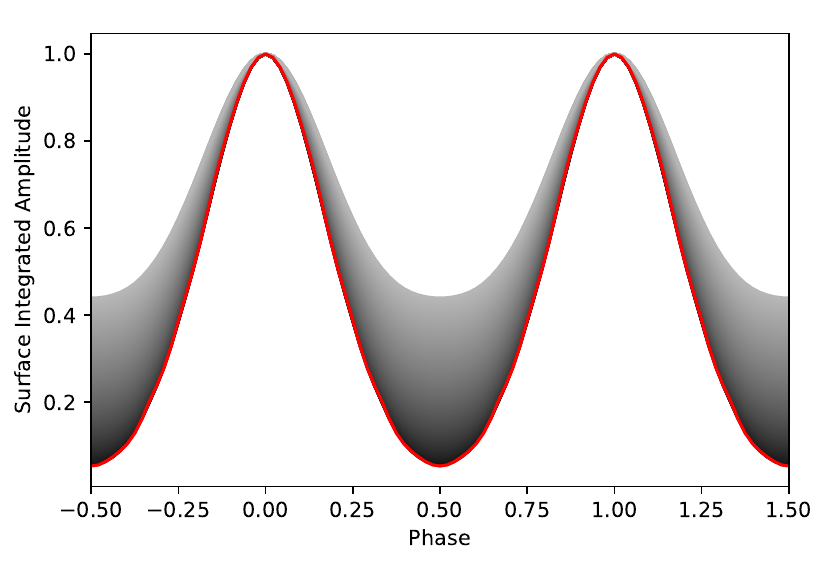} \hfill
    \includegraphics[trim=0 238 970 258, clip=true, width=.32\textwidth]{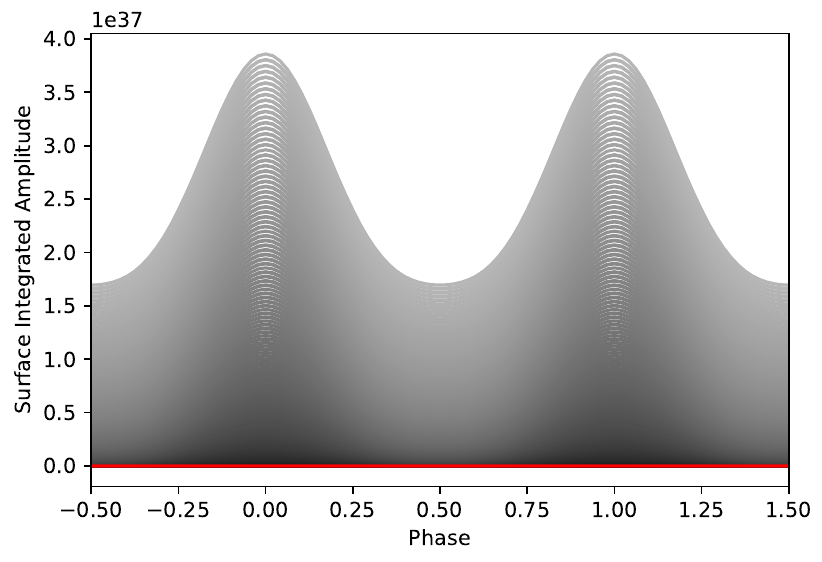} \hfill
    \includegraphics[trim=0 0 0 20, clip=true, width=.32\textwidth]{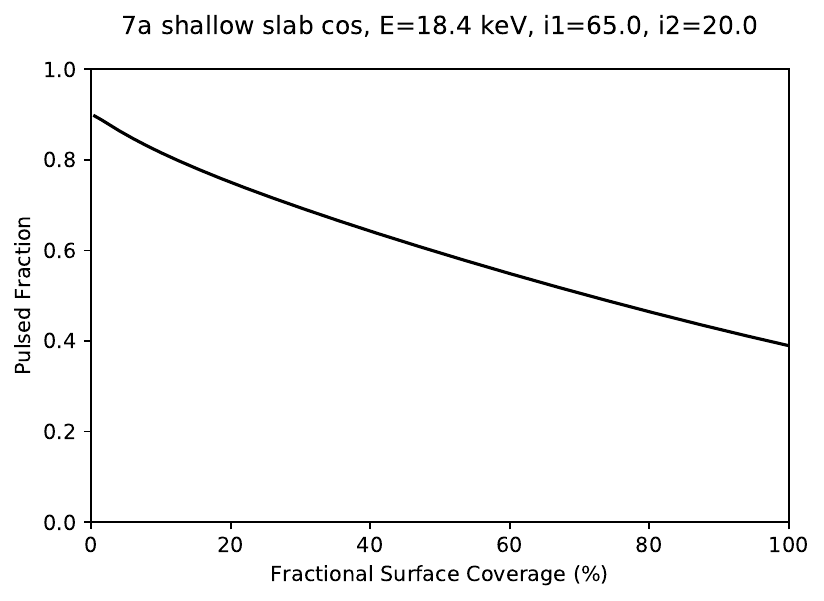}  
    \caption{The Centaurus X-3 Phantom Figures (See Appendix) along with the corresponding variation in the pulsed fraction. (\textit{left}) Normalized and (\textit{middle}) non-normalized surface-integrated pulse profiles are isochors with a $0.5\%$ change in the fractional area coverage for known pulsar inclinations $(i_1=65\textsuperscript{o}, i_2=20\textsuperscript{o})$ for a shallow slab beam for 15.2 keV without light bending. Lighter shade denotes greater surface coverage over the full $0-100\%$ range. The red curve depicts a standard model hotspot with an area of 1 km$^2$. The flux is of the order of 10$^{37}$ erg/s which is typical of accreting X-ray pulsar luminosities. (\textit{right}) The corresponding variation in pulsed fraction with fractional surface coverage.}
    \label{fig:varyA_slab_65_20}
\end{figure}

Various simulated pulse profiles for Centaurus X-3 are shown in Figs. \ref{model_simu}, \ref{fig:varyA_slab_65_20} and \ref{fig:vary_M2}.
Given a pencil beam geometry for low luminosity (See Table \ref{tab:lumi} for details) and the known values of inclinations as ($i_1=65\textsuperscript{o},~i_2=20\textsuperscript{o}$), it would not be possible for the beamed radiation to present a prominent secondary inter-pulse in the Centaurus X-3 profile, as the viewing angles sampled by the observer's line-of-sight range from $i_1+i_2 = 85\textsuperscript{o}$  and $|i_1-i_2| = 45\textsuperscript{o}$. The secondary inter-peak which vanishes for higher energy X-rays is known to be due to a distorted magnetic dipole configuration \citep{1996ApJ...467..794K, 2008ApJ...675.1487S}. \cite{1996ApJ...467..794K} had proposed an offset of 10\textsuperscript{o} of the emission regions from the dipole axis. A value of ($49\textsuperscript{o}, 17\textsuperscript{o}$) has been suggested recently by \cite{2022ApJ...941L..14T} using polarimetric characterization with \textit{IXPE}.

The simulated pencil profiles in Fig. \ref{model_simu} seem to resemble the Cen X-3 profile well\footnote{Centaurus X-3 has a mixed beam which can be in the pencil emission mode for low luminosity states or in the fan emission mode for high accretion rates.} confirming a low luminosity state. Furthermore, all the lower energy profiles exhibit a miniscule secondary inter-pulse typically seen in the Cen X-3 profile.

\subsection{Observed AstroSat/LAXPC pulsed profiles for Centaurus X-3}
The pulse profiles observed with \textit{AstroSat}/LAXPC of Centaurus X-3 are seen in Fig. 3 of \cite{2021JApA...42...58S}. The pulse profiles are derived using time-of-arrival analysis by modeling the Doppler modulation arising from the orbital motion of stars in a binary system. The folded broad-band 3.0 -- 80.0 keV pulse profile phase-locked in Fig. 2 of \cite{2021JApA...42...58S} -- was resolved into four different equi-photon energy bands viz. 3.0 -- 6.0 keV, 6.0 -- 9.0 keV, 9.0 -- 15.0 keV and 15.0 -- 40.0 keV as seen in Fig. 3 of \cite{2021JApA...42...58S} after correcting for essential effects for space-based measurements \citep{2021JApA...42...58S}. The 40.0 -- 80.0 keV data was discarded as the flux in this band was found to be negligibly small during the observation span. The energy band 15.0 -- 40.0 keV was not resolved further as the pulse shape does not seem to change within this energy range, as seen in Fig. 3 in \cite{bachhar2022timing} for the 15.0 -- 22.0 keV and 22.0 -- 35.0 keV profiles for an extended \textit{AstroSat} data-set.

Note that the \textit{X}-axis of the observed energy-resolved profiles (Fig. 3 of \cite{2021JApA...42...58S}) is fiducial. As per convention, the profiles are calibrated such that the plot window begins with an increasing trend in flux, with the pulse minimum marked as spin phase \texttt{0.0}. In reality, though, the pulse peak will arise from spin phase \texttt{0.0} for a pencil beam and \texttt{0.25} for a fan beam -- but the observational degeneracy between the two models remains to be resolved -- hence, the aforementioned plotting convention. On the other hand, the \textit{X}-axis of the simulated profiles is well-calibrated and maps the pulsar spin phase to the corresponding flux value in the pulse profile.

\begin{table}
    \centering
    \caption{Pulsed fractions for the energy-resolved pulse profiles of Centaurus X-3 in Fig. 3 of  \citep{2021JApA...42...58S}. }
    \label{tab:PF}
    \begin{tabular}{ll}
    \toprule \hline
        Energy band & Pulsed Fraction \\
        & \hspace{.7cm}(PF) \\
        \hspace{.4cm}(keV) & \hspace{.7cm}($\%$) \\
        \midrule
        3.0 -- 6.0 &  51.8 $\pm$ 0.34\\
        6.0 -- 9.0 & 53.6 $\pm$ 0.37 \\
        9.0 -- 15.0 & 58.2 $\pm$ 0.35 \\
        15.0 -- 40.0 & 48.4 $\pm$ 0.43 \\
        \bottomrule \hline
    \end{tabular}
\end{table}

\section{Spherical Caps}

To allow for larger hotspot radii in the following sections for a deeper comparative study between the simulated and observed results, the hotspot is allowed to span larger areas on the surface of the neutron star. The polar co-ordinate system does not fare well with this choice and its well-suited approximation to smaller polar caps breaks down. The deviation appears in the form of a sinc function (See Eq. (\ref{sinc})) at larger distances from the magnetic pole. The polar co-ordinate geometry must be transformed into a spherical co-ordinate geometry for performing the double integration over the closed hotspot surfaces. The polar area element will be modified into a spherical area element, which continues to lie flush on a larger extent of the neutron star surface.

\begin{figure}
    \centering 
    \includegraphics[clip, trim = 0 0 130 0, width=\textwidth]{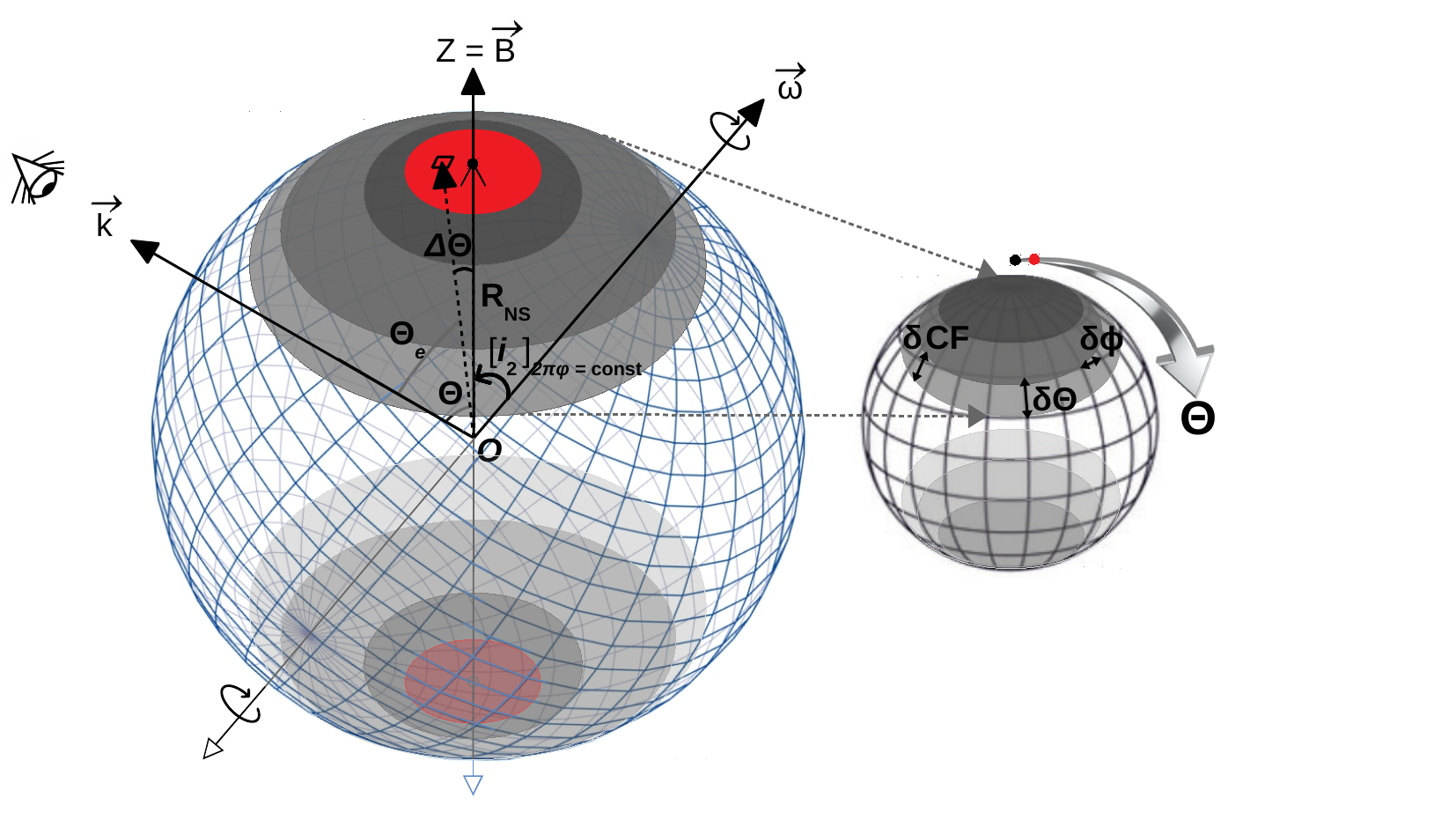}
    \caption{Sketch of the geometric construction of a finite spherical grid $(\phi, \Theta)$ on model antipodal spherical slabs lying at the polar caps of a pulsar with varying radii shown by fainter gray contours. The standard emission area of 1 km$^2$ is shown in red colour. The overall pulsar geometry is characterized by the observer's line of sight $\vec{k}$, magnetic axis $\vec{B}$ and the rotation axis $\vec{\omega}$ and the corresponding angles are the same as shown in Figs. $1-3$ in the accompanying paper, where the $Z$-axis is rotated by an angle $i_2$ keeping the value of the spin phase constant. $\Delta \Theta$ is the slab-resolved polar angle of an arbitrary infinitesimally small spherical area element and $R_{\text{NS}}$ is the radius of the neutron star (= 10 km, for a model pulsar). \textit{(inset to the right)} $(\delta \phi, \delta \Theta)$ represent the spherical grid cell resolution.}
    \label{fig:diagram_sph_int}
\end{figure}

\subsection{Co-ordinate transformation to spherical geometry for the construction of a curved discretized grid} \label{spherical}
While $\Delta A = r dr d\phi$, with $r$ being the radial co-ordinate and $\theta$ being the azimuthal co-ordinate is a good approximation for small hotspot radii ($r<<R_{\text{NS}}$$\sim$$10$) km like a model hotspot with a standard area of 1 km$^2$ (\cite{MeszarosNagel1985b}; using the Stefan-Boltzmann law). However, for areas with a larger emission span that can cover greater fractional surface areas over the full range of $0-100\%$ coverage of the neutron star surface, spherical co-ordinates must be used for exact results as shown in Fig. \ref{fig:diagram_sph_int}. The formula for surface integration \citep{Dipanjan2012} changes to $\Delta A = R^2_{\text{NS}} \sin\Theta d\Theta d\phi$, i.e.

\begin{equation} \label{sph}
    F = \sum_{\phi} \sum_{\Theta} [I(\theta)\cos{\theta}] \big(R^2_{\text{NS}} \sin{\Theta} d\Theta d\phi \big)
\end{equation}

The existing polar co-ordinate framework is transformed into a spherical structure\footnote{\label{cyl} Similarly, a cylindrical co-ordinate system $(\rho, \phi, z)$ for an accretion column will require the formula to be changed to,
\begin{equation}
    F =  \sum_{\phi} \sum_{z} [I(\theta)\cos{\theta}] \Delta A(\phi, z) = \sum_{\phi} \sum_{z} [I(\theta)\cos{\theta}] (\rho d\phi dz).
\end{equation}} as,
\begin{equation}
    F = \sum_{r} \sum_{\phi} [I(\theta)\cos{\theta}] \times R_{\text{NS}} dr d\phi \sin\Bigg(\frac{r}{R_{\text{NS}}}\Bigg) = \sum_{r} \sum_{\phi} [I(\theta)\cos{\theta}] (r dr d\phi) \times \frac{\sin(r/R_{\text{NS}})}{(r/R_{\text{NS}})}~.\label{sinc} 
\end{equation}
where, the sinc function $\frac{\sin(r/R_{\text{NS}})}{(r/R_{\text{NS}})}$ is the correction factor for a radial co-ordinate $r$ measured along the spherical surface of the neutron star within the hotspot. Although the use of this spherical correction suffices for computational purposes, the original flux formula in Eq. (\ref{sph}) is used in this work for exact results.

The fractional surface coverage is varied over the full range [0, 100\%] at a fine resolution of 0.5\% in Figs. \ref{fig:varyA_slab_65_20}, \ref{phantom} and \ref{fig:intercepts}.  In Figs. \ref{fig:varyA_slab_65_20} and \ref{phantom}, black colour is used to indicate the 0\%~i.e. \lq point' hotspots case and red colour is used for model circular slabs with a radius of 0.564 km corresponding to the standard emission area of 1 km$^2$. Increasingly fainter grayscale curves depict hotspots with larger radii and surface coverage. 

\section{Fractional surface coverage}
The fraction of the compact star's surface occupied by the emission region is varied in this section. In the case of a hotspot, this is given by the fractional area of two flat, antipodal circles lying flush on a spherical neutron star surface. Since a spherical co-ordinate system is used, it is desirable to represent these in terms of the radius of the neutron star $R_{\text{NS}}$ (model value = 10 km), the azimuthal angle $\phi$ = [0, 2$\pi$] and the polar angle $\Theta$ = [0, $\pi$]. The circumference of the hotspot is parameterized by the polar half angle $\phi$ measured from the direction of the magnetic pole of the neutron star ($\phi_p$ = 0\textsuperscript{o}). 

From trigonometric considerations for the spherical cap of a compact star, the standard formula for the area of a finite region $A$ lying flush on the surface of the neutron star is given as,
\begin{equation} \label{h}
    A  = \int_{\theta}^{\theta'}\int_{\phi}^{\phi'} R^2 \sin{\Theta} d\Theta d\phi = R^2 (\phi' - \phi) ({\cos{\Theta} - \cos{\Theta'}}).
\end{equation}
Note that there are two antipodal hotspots, so, the area of a single hotspot fractionally covers one hemisphere of the neutron star while two hotspots span the total surface area, using symmetry arguments under the assumption of identical emission areas. Thus, the corresponding fractional coverage by the hotspot is the ratio of the area covered by both the hotspots to the full surface area of the neutron star or the area covered by a single hotspot to the area of a single hemisphere of the neutron star. Using Eq. (\ref{h}), the coverage fraction ($CF$)
\begin{equation} 
    CF = \frac{R^2(2\pi) (\cos{\Theta(0)} - \cos{\Theta'})}{2\pi R^2}
\end{equation}
Thus, the percentage surface coverage of the neutron star is calculated as,
\begin{equation} \label{CF}
    CF ~(\%) = (1-\cos{\Theta'}) \times 100,
\end{equation}
where $\Theta'$ denotes the polar (half) angle corresponding to the hotspot radius, with unprimed and primed quantities representing lower and upper limits of integration i.e. the polar (half) angle corresponding to the circumference of the spherical cap measured from the center of the neutron star.

\section{Pulsed fraction as an Estimator} \label{sec:pf} 

The pulsed fraction ($PF$) is proposed as a robust quantitative measure for comparison between the simulations developed in the accompanying paper with the corresponding observations of binary accreting X-ray pulsars. The pulsed fraction is a physical quantity that measures the fractional variability in a signal i.e. the varying component in a pulsar profile after removing the continuum component, down-weighted by the overall flux received. It is a quantitative characteristic of the pulse shape depending only on the maximum and minimum flux values, and therefore more robust against stochastic shape fluctuations. It is the difference between the maximum ($f_{max}$) and minimum ($f_{min}$) flux values normalized by their sum and is given as $PF$ = $(f_{\text{max}}-f_{\text{min}})/(f_{\text{max}}+f_{\text{min}})$. It represents the relation between the emission coming from the emission region seen as the \lq pulsed' component in comparison to other persistent emissions received as the \lq non-pulsed' component. As the emission functions are continuous (Caveat: there are discontinuities observed near the cyclotron energy), the extremum flux values typically correspond to the inclination extrema and thus, it is the angular sweep $(i_1+i_2)$ and $|i_1-i_2|$ that matters. For flat profiles, $f_{max}=f_{min}$ as seen in Fig. \ref{fig:vary_M2} and the numerator vanishes. 
Isotropic slabs with a cosine factor seen in Fig. \ref{dracula_fangs} provide broader pulse profiles compared to \citeauthor{MeszarosNagel1985b}'s beams.

\section{Results and Discussions} \label{sec:rnd}

\subsection{Comparison with observations}

The required observed Centaurus X-3 pulse profiles derived from the Large Area X-ray Proportional Counter (LAXPC) on-board \textit{AstroSat} were reported in \cite{2021JApA...42...58S} (See \cite{bachhar2022timing} for an extended analysis). The required simulated pulse profiles for comparison with the observed profiles of Centaurus X-3 are shown in Fig. \ref{model_simu}. With a reasonable assumption that the hotspot geometry is governed by the interplay of the accretion rate and X-ray luminosity, and is not energy-dependent\footnote{In the unlikely case that it is, one would have to give way to considering energy-resolved polar cap area distribution and spectrally-weighted hotspot area calculations. The energy-dependence of the pulsed fraction is discussed in Sec. \ref{depend}.}, any suitable energy band among the energy-resolved pulse profiles of any arbitrary pulsar available in the literature and lying close to the available set of discrete values used by the \texttt{AXP4} is sufficient for practical comparison. This way, the work in this paper is easily extendible to multiple pulsars available in known galactic catalogs with an eye out for individual deviations from model behaviour on a case-to-case basis.

We choose 15.2 keV (fourth row from bottom in Fig. \ref{model_simu}) as it not only lies within the 15.0 -- 40.0 keV band (with complete pulse shape stability throughout) but also in the 15.0 -- 22.0 keV band of \cite{bachhar2022timing} (referred later in Sec. \ref{depend}) as well as near the center of the 3.0 -- 80.0 keV broad-band profiles (given that emission flux in the 40.0 -- 80.0 keV range was found to have negligibly small emission). Most importantly, it does not exhibit a secondary inter-pulse which arises in all the other bands due to a deviation from the assumed perfectly dipolar configuration\footnote{gravitationally bent additional flux seen from the unseen face of Cen X-3 could also be an additional contributing factor} \citep{1996ApJ...467..794K} that went into the numerical simulations for the purpose of generalization. 

\begin{figure}
    \centering
    \includegraphics[width=0.6\linewidth]{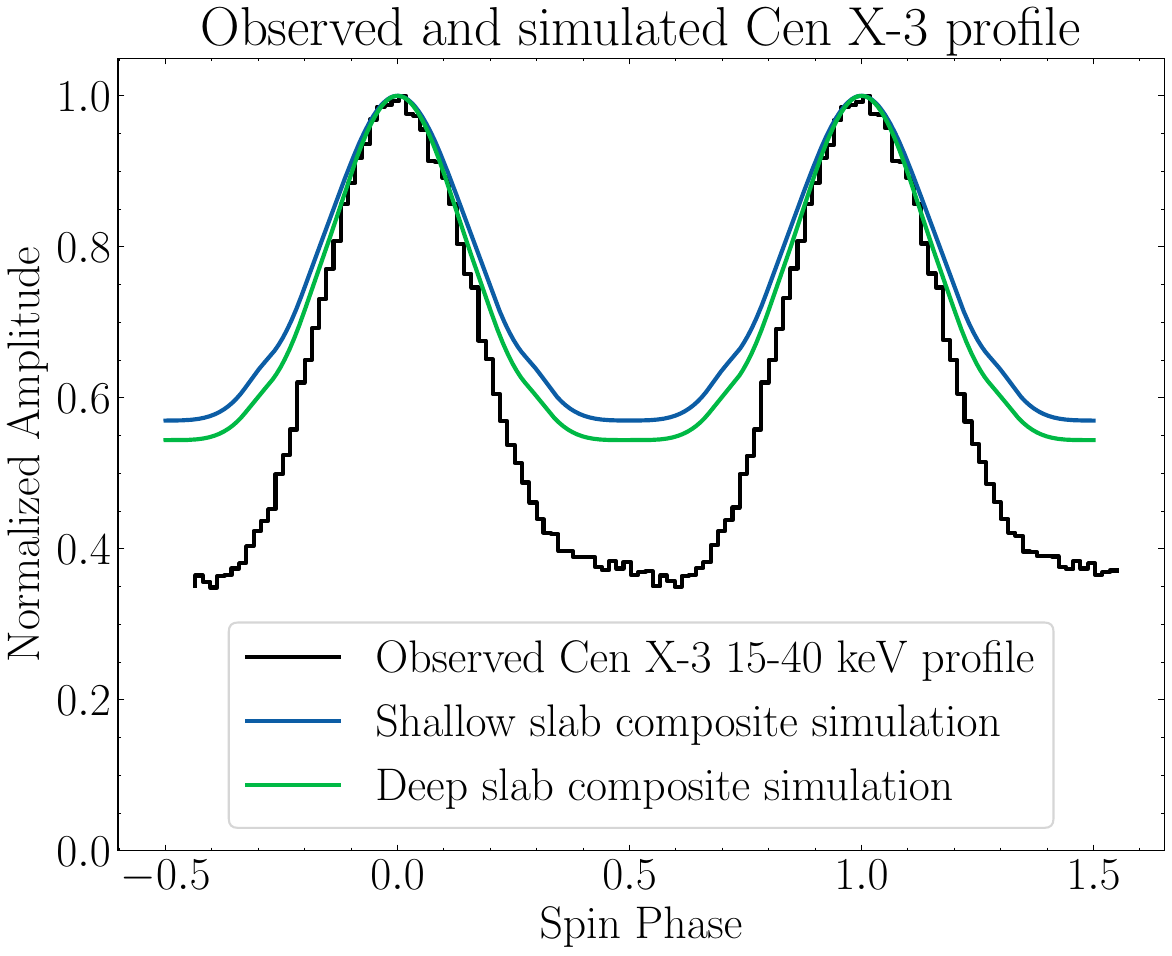}
    \caption{\textit{(Black)} Observed \textit{AstroSat}/LAXPC 15.0 -- 40.0 keV Centaurus X-3 pulse profile over-plotted with the composite gravitationally bent and slab-integrated simulated profiles for (65\textsuperscript{o},~20\textsuperscript{o}) inclination at 15.2 keV for two antipodal \textit{(blue)} shallow and \textit{(green)} deep slabs of standard 1 km$^2$ at \texttt{0.001} spin phase resolution. As the $X$-axis of the observed Cen X-3 pulse profiles in Fig. 3 of \cite{2021JApA...42...58S} is fiducial (marked with pulse minimum at phase \texttt{0.0} as per convention) due to the pencil and fan beam degeneracy, this profile is suitably shifted along the phase \textit{X}-axis to coincide their pulse maxima with the simulated pulse peaks. (Plot credit: Mr. Swarnim Shirke)}
    \label{fig:compare}
\end{figure}

The observed and simulated slab profiles are over-plotted in Fig. \ref{fig:compare}. It can be seen that the overall profile is sufficiently reproduced with the deeper optical depth for emission geometries being a little closer to the observed profile. However, since the exact value of optical depth that can match the profile is unknown and remains to be seen if practical, I keep to the shallow case as a model geometry allowing for examining the full span of the half-angle made by the outer circumference of the hotspot with the magnetic axis. 

\subsection{Constraining the polar cap radius}

This paper improves upon previous studies of the changes in the pulse profile with the area of the hotspot by performing high-resolution simulations for the polar half angle. It allows for examining the full span ($0-100\%$) of the half-angle made by the outer circumference of the hotspot\footnote{This assumes a circular slab as in \cite{MeszarosNagel1985a}. There are other different possible beaming configurations, namely, (i) isotropic emitter, (ii) isotropic emitting slab geometry, (iii) cylindrical column \citep{MeszarosNagel1985a}, (iv) hollow column \citep{2001ApJ...563..289K}, (v) conical column and (vi) mound \citep{Dipanjan2012, 2020MNRAS.497.1029B}, (vii) multiple oval spots \citep{2019ApJ...887L..21R, miller2019psr} and (viii) filled crescents \citep{miller2019psr}. Some of these alternatives are discussed to a limited extent in the accompanying paper.} with the magnetic axis. 
As summarized in Table \ref{tab:GBL}, the estimated area of the hotspot is seen to drop by two orders of C.G.S. magnitude upon accounting for the effect of gravitational bending of light, falling within the ballpark of the model emission area of $10^{10}$ cm$^2$ (See Fig. \ref{fig:intercepts} for inferential details).

\subsection{Gravity-dependence of the pulse profile} \label{g}
As seen in Fig. \ref{fig:intercepts}, there is a substantial suppression in pulsed fraction upon accounting for general relativistic effects. This is directly visible in the simulated profiles in Figs. 6 and 7 of the accompanying paper. As shown in Table \ref{tab:GBL}, the coverage fraction is $\approx$$75\%$\footnote{The discussions of \cite{1981A&A...102...97W}, do consider cases of very large polar caps (with polar half angle $>30$\textsuperscript{o}) as being physically realistic.} of the neutron star area without GR and reduces by an order of magnitude to $\approx$$9\%$ with light bending. The case is similar for the polar cap half-angle. The hotspot area for Centaurus X-3 using the numerical simulations is $\approx$$5 \times 10^{12}$ cm$^2$. Upon accounting for the effect of gravitational bending of light, it seems to drop by two orders of magnitude within the ballpark of the standard model value of $\approx$$10^{10}$ cm$^2$. The hotspot radius itself drops from $12.27$ km (which remains practical given the high surface coverage in view of recent values of $12.45\pm0.65$ km reported by \cite{miller2021radius} for the radius of a model 1.4 M$_{\odot}$ neutron star) to $1.36^{+0.29}_{-0.26}$ km, the latter subsuming the model value of 0.564 km for a 1 km$^2$ circular slab. This is especially impressive given that the curve in Fig. \ref{fig:intercepts} is highly sensitive to the value of the pulsed fraction, such that even a small change in the pulsed fraction corresponds to a large change in the fractional surface coverage. The errors in the pulsed fraction in Table \ref{tab:PF} are marked in the horizontal line shown in Fig. \ref{fig:intercepts} and are propagated further through their regions of intersections with the simulated curves for a full range of fractional surface coverage.
\begin{table}
\begin{threeparttable} 
    \caption{Surface parameters of Centaurus X-3 inferred from simulations using pulsed fraction as a quantitative estimator (\textit{fifth column}) excluding and (\textit{last column}) including the effect of gravitational bending of light (GBL). See Fig. \ref{fig:intercepts} for inferential details.}
    \label{tab:GBL}
    \begin{tabular}{lcccll}
    \toprule \hline
      Physical Quantity  & Notation & Unit & Trigonometric Formula &  Without GBL & With GBL\\
    \midrule
      Coverage Fraction  & $CF$ & (\%) & from Fig. \ref{fig:intercepts} & $75.3 \pm 0.1$ & $9.3^{+3.9}_{-3.5} $ \\
      Polar half-Angle & $PA$ & ($\textsuperscript{o}$) & 
      $PA = \cos^{-1}(1-CF)$~\tnote{a} & $75.7 \pm 2.56$ & $7.82^{5.06}_{4.98}$\\
      Hotspot Area & $A$ & ($\times 10^{10}$ cm$^2$)& $A$ = $PA$ (radian) $ \times~2\pi R^2_{\text{NS}}$~\tnote{b}  & $473.1 \pm 0.6$ & $5.84^{2.45}_{2.2} $\\
      Hotspot radius & $r$ & (km) & $r=\sqrt{A/\pi}$~\tnote{c} & $12.27$~\tnote{d} & $1.36^{+0.29}_{-0.26}$ \\
      \bottomrule \hline
    \end{tabular}
        \begin{tablenotes}
    \item[a] Inverting Eq. (\ref{CF}) for the neutron star as a sphere.
    \item[b] For a hemisphere governed by the magnetic axis of the neutron star.
    \item[c] Circular slab approximation.
    \item[d] Insignificant propagated error = $\pm 0.0000778$ km.
    \end{tablenotes}
    \end{threeparttable}
\end{table}
\subsection{Luminosity-dependence of the pulsed fraction} \label{depend}

\begin{figure}
    \centering
    \includegraphics[height=.43\textheight, clip=true, trim = 0 0 50 43]{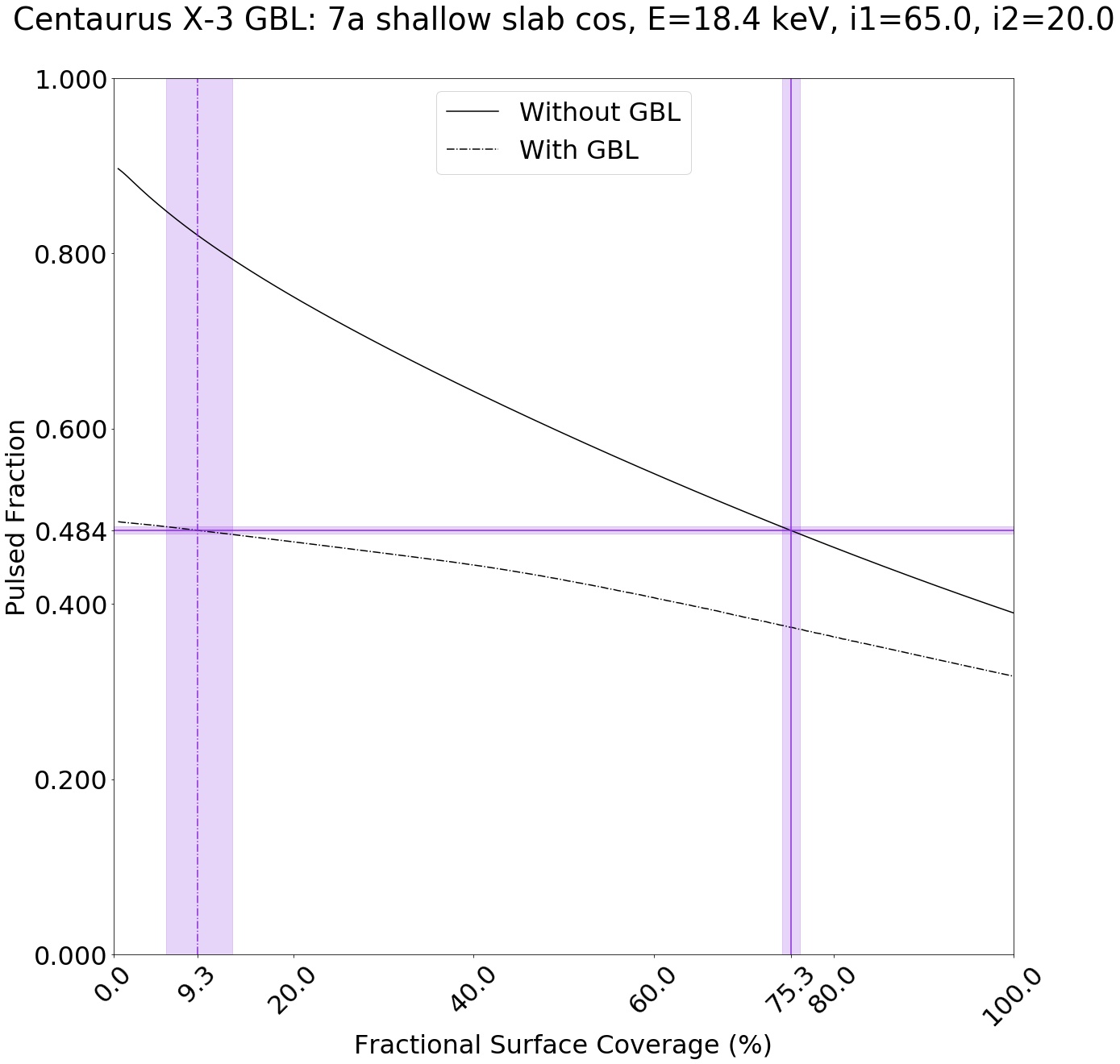}\\
    \includegraphics[height=.43\textheight, clip=true, trim = 0 0 50 43]{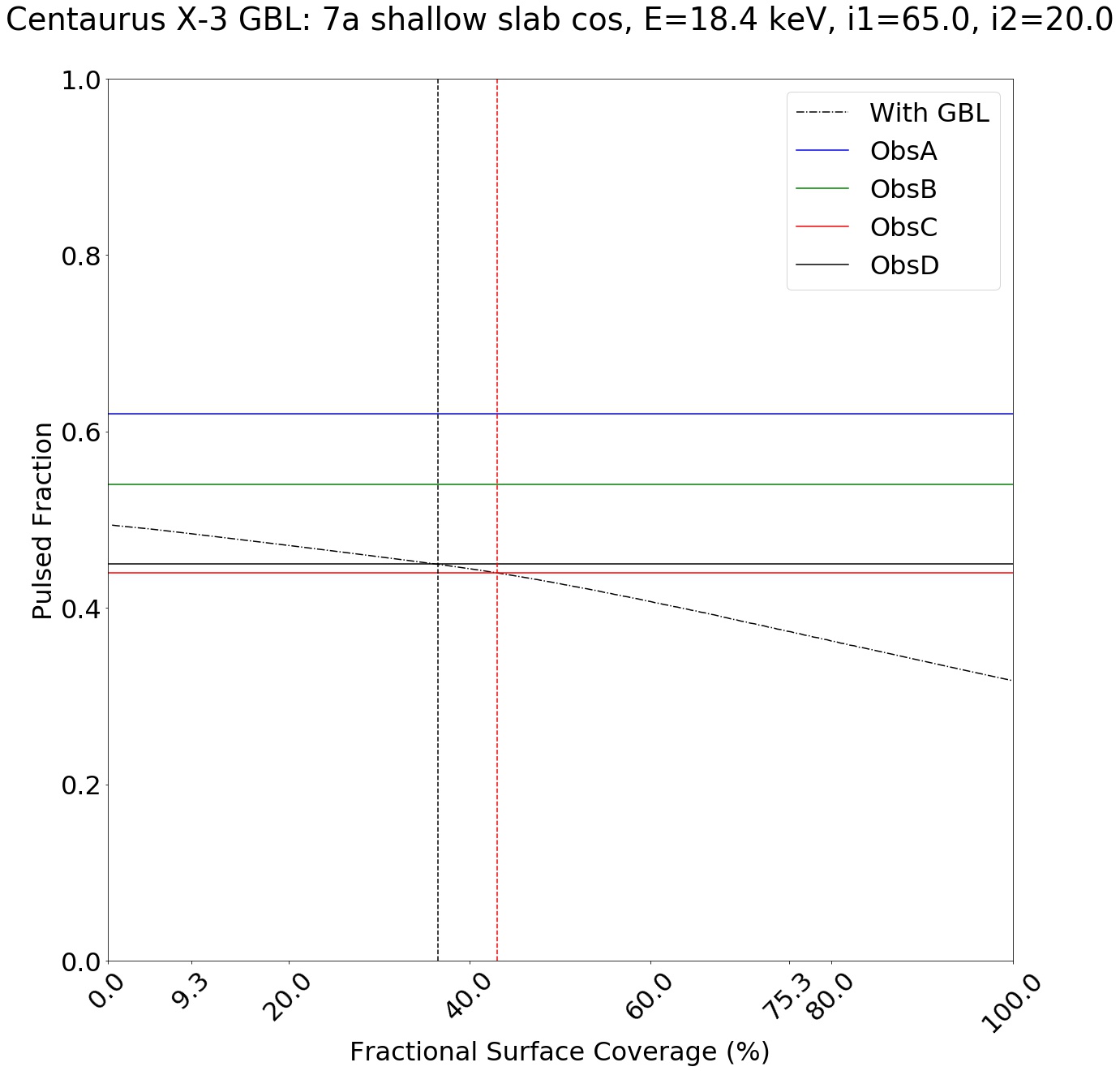}
    \caption{Relation between pulsed fraction and fractional surface coverage for the Centaurus X-3 pulsar. (\textit{top}) Variation in the pulsed fraction with the fractional surface coverage \textit{(solid curve)} with and \textit{(dashed curve)} without including the effect of composite gravitational bending of light for the Centaurus X-3 (65\textsuperscript{o}, 20\textsuperscript{o}) inclination at $15.2$ keV. The curves are obtained at a 0.5\% resolution of surface coverage percentage. Abscissae with error lines are located for the points of intersection of (\textit{horizontal line}) the pulsed fraction (with error lines) of the 15.0 -- 40.0 keV pulse profile (Refer Table \ref{tab:PF}). The variation due to gravitational bending is far higher compared to that produced by a variation in the surface coverage thus producing non-intersecting curves and preventing the occurrence of possible degeneracies in the corresponding results. (\textit{bottom}) Composite gravitational bending-included luminosity-dependence of the fractional surface coverage through the pulsed fraction using the $15.0 - 22.0$ keV profiles from \cite{bachhar2022timing} (See Table \ref{tab:lumi} for further details). The colors used for the four different values of luminosity are kept consistent with \citeauthor{bachhar2022timing}'s choice. Values of abscissae are obtained using vertical line construction at the points of intersection with observed pulsed fractions shown in horizontal lines across the $Y$-axis.}
    \label{fig:intercepts}
\end{figure}

We use the profiles reported by \cite{bachhar2022timing} for different luminosity states\footnote{\url{http://www.graphreader.com/}. See also, a similar \lq Graph2Data' package built using an algorithm suggested by the author and made available in a GitHub repository by Dr. V. Upendran: \url{https://github.com/Vishal-Upendran/Graph2Data.git}.} for examining the effect of luminosity on the pulsed fraction (Refer Fig. 6 in \cite{2022arXiv220414185M} for the apparent luminosity and luminosity range observed in some X-ray transients powered by accretion onto magnetized NSs). The color scheme namely, blue for ObsA, green for ObsB, red for ObsC and black for ObsD respectively, is retained for consistency. Firstly, the high flux observations do not intersect the gravitationally bent simulated curve for the pencil beam in Fig. \ref{fig:intercepts}. Owing to the high luminosity, this might be attributable to a switch to a fan-beam emission as Centaurus X-3 is known to exhibit a mixed pencil-fan nature. The lower luminosity curves do intersect the slab geometry curve -- consistent with the pencil beam expectation -- with a decrease in pulsed fraction and corresponding increasing surface coverage with a reduction in luminosity which is consistent with the known explanation for the formation of a magnetically confined column of accreted plasma. This points to lower accretion rates ($\dot{M}$) giving rise to slabs and higher accretion rates corresponding to columns that can survive super-Eddington regimes. With the slab regime, lower $\dot{M}$ corresponds to a wider accretion funnel that strikes the neutron star surface allowing for bigger hotspots and lower pulsed fraction. Bear in mind that the emission from columns is broad because of a \lq fan' beam in which X-ray photons escape from the \textit{side} walls of the column. %(Note that the broader profiles for columns result in an increase in emission area along the column height, not radius, at constant emission area size.)

\begin{figure}
    \centering
    \includegraphics[width=.45\textwidth, clip=true, trim=0 0 350 30]{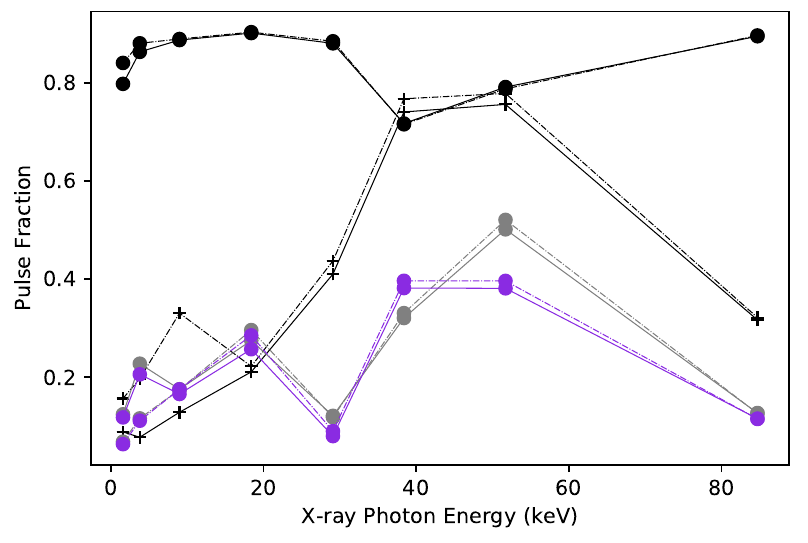} 
    \includegraphics[width=.5\textwidth, clip=true, trim=0 0 0 30]{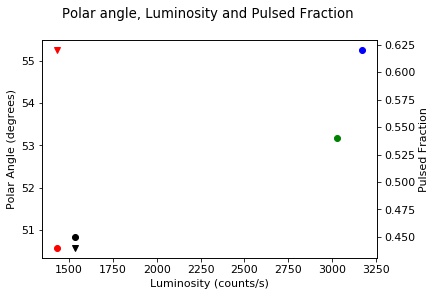}
    \caption{(\textit{left}) High-resolution energy-dependence of the simulated pulsed fraction for slabs (o) and columns (+) for (\textit{solid lines}) shallow and (\textit{broken lines}) deep geometries corresponding to the Centaurus X-3 (65\textsuperscript{o}, 20\textsuperscript{o}) inclination. Black curves are for model, beamed profiles, gray after gravitational bending and violet for surface-integrated slabs. (\textit{right}) Luminosity-dependence of the observed pulsed fraction ($PF$) and estimated polar half-angle ($PA$) of circular hotspots (by comparison with pencil beam simulations). The colours are the same as in \cite{bachhar2022timing}. Dots represent pulsed fractions shown on the right \textit{Y}-axis and inverted triangles denote the corresponding polar angle measurements from Fig. \ref{fig:intercepts} shown on the left \textit{Y}-axis.}
    \label{fig:dependence}
\end{figure}

The right panel in Fig. \ref{fig:dependence} exhibits a clear positive correlation between the pulsed fraction and the X-ray luminosity in the pencil regime. Considering that the lower luminosity values correspond to slabs, a decrease in the composite polar angle is observed with an increase in X-ray luminosity. Table \ref{tab:lumi} presents the corresponding source parameters using this luminosity-dependence of the polar half-angle. In keeping with the trend of the polar half-angle calculated using the composite gravitationally bent and surface-integrated pulse profiles, the coverage fraction also decreases with an increase in luminosity from ObsC to ObsD i.e. brighter pulsars in the low luminosity regime have a higher pulsed fraction, and therefore, more collimation. They correspond to a smaller hotspot area and more efficient funneling of the accreted matter onto the polar cap accompanied by a narrower outgoing X-ray beam. The corresponding values of the hotspot area exhibit a decrease but remain within the same order of magnitude, unlike in Sec. \ref{g}. In this exercise, it is the model dependence or trend that is important as a proof-of-concept -- the actual values could improve on considering a deeper geometry (See Fig. \ref{fig:compare}). As the luminosity continues to increase, it would lead to a transition to the fan state through the formation of an accretion column. A sudden turn-over in pulsed fraction would arise from a change from pencil to fan emission mechanism. The height of the resultant column ($z$) would be an additional parameter that could determine further luminosity-dependence of the pulsed fraction\footref{cyl}.

\begin{table}
    \centering
    \caption{Luminosity-dependence of the source parameter estimates for four different luminosity states as per \cite{bachhar2022timing}.}
    \label{tab:lumi}
    \begin{tabular}{lccccll}
    \toprule \hline
      Observation & Notation & Unit &Blue & Green & Red & Black \\
        & & & A & B & \hspace{.3cm}C & \hspace{.3cm}D \\
        Flux & $f$ & (counts/s) & 3170 & 3029 & 1432 & 1534 \\
        \midrule
       Pulsed Fraction & $PF$ & & 0.62 & 0.54 & 0.44 & 0.45 \\
       Coverage Fraction & $CF$ & (\%) & -- & -- & 43\% & 36.5\% \\
       Composite Polar half-Angle & $PA$ & (\textsuperscript{o})& -- & -- & 55.25 & 50.58 \\
       Hotspot area & $A$ & ($\times 10^{12}$ cm$^2$) & -- & -- & $2.7$ & $2.29$ \\
       Hotspot radius & $r$ & (km) & -- & -- & 9.3 & 8.54 \\
       \bottomrule \hline
    \end{tabular}
\end{table}

\subsection{Energy-dependence of the pulsed fraction} \label{energy}

The top panel of Fig. \ref{fig:dependence} re-affirms a marked reduction in the pulsed fraction with general relativistic and surface integration effects. An increasing optical depth seems to result in higher pulsed fractions across all cases with the difference being larger for columns. In the case of slabs, it grows more pronounced upon the inclusion of GR effects and surface emission. The noteworthy feature seems to be around the cyclotron line energy $\sim$38 keV in the 25.0 -- 40.0 keV band. While the model beamed profiles exhibit a decrease followed by an increase in emission within this band -- as opposed to the inverted trend in the column -- both the corrected profiles, after accounting for the important physical effects that modify their observed behaviour, exhibit an increase in the pulsed fraction in this band. Perhaps, this dip feature in the energy-resolved pulsed fraction trend could suffice as a purely theoretical alternative to polarimetry in breaking the existing degeneracy between the pencil and fan beam models.

\section{Conclusions} \label{sec:conclusions}
Known parameters for Centaurus X-3, a promising target for X-ray observations are provided as input to the numerical code \texttt{AXP4} developed in the accompanying paper (Part -- I) and the resultant simulated pulse profiles are compared with the observed $15.0 - 40.0$ keV pulse profiles derived from \textit{AstroSat}/LAXPC in \cite{2021JApA...42...58S}. Extra energy bands that bear signatures of deviation from the assumed behaviour are omitted in this exercise. The profile fit favours a deeper optical depth of the emission region. The variation in surface coverage by the polar cap to very large spherical caps (with polar half-angle $>30$\textsuperscript{o}) is permitted by building a new, custom \lq \texttt{Surface Coverage}' numerical module (See Appendix) which can handle the full range of percentage coverage ($0-100\%$).  The pulsed fraction is chosen as a robust estimator to compare the simulated and observed profiles. 

The hotspot radius for Centaurus X-3 is inferred to be $12.27$ km, which drops by an order of magnitude to a realistic $1.36^{+0.29}_{-0.26}$ km (within the ballpark of the standard model value of $\sim$1 km) on accounting for the effect of gravitational bending of light. General relativistic bending is seen to produce a first-order modification that dominates the second-order variation due to varying surface coverage without the presence of any degeneracies. As the pulse profile is sensitive to luminosity variations, the correlation of the size of a finite polar cap and its dependence on X-ray luminosity -- through the rate and subsequently, the geometry of accretion -- demonstrates a pencil beam for low luminosity states and favours a fan beam for high luminosity states in the $15.0 - 22.0$ keV band.

Besides the energy-dependent dip feature discussed in the previous Sec. \ref{energy}, the effect of gravitational light bending and/or variation with the area of emission region through the compactness parameter $u$ may provide another purely theoretical alternative to X-ray polarisation studies for breaking the existing pencil beam and fan beam degeneracy. Although a single, model pulsar was chosen for this thesis, to demonstrate the depth of physical and astrophysical prospects of such a study, the \texttt{AXP4} code has been generalized for application to a wide range of cases from known pulsar catalogs, especially, with the possible inclusion of accretion columns (using cylindrical co-ordinate transformation) in the future. 

\begin{acknowledgments}
Discussions with Dr. Y. Bhargava are acknowledged. Prof. R. Srianand and Prof. R. Misra provided chronological facilitation for formally reporting the work as a paper, supported by Prof. A. N. Ramaprakash. %\textcolor{gray}{Constructive comments and suggestions from an anonymous referee have improved the content of the paper.}
\end{acknowledgments}

\facilities{Dell OptiPlex 5060 Desktop\footnote{\url{https://www.dell.com/learn/in/en/inbsd1/shared-content~data-sheets~en/documents~optiplex_5060_spec_sheet.pdf}} with an Intel\textsuperscript{\textregistered} Core\textsuperscript{TM} i7-8700T processor, Ubuntu\textsuperscript{\textregistered} 18.04.3 LTS (64-bit) Operating System, 8 GB DDR4 RAM \& 2 TB HDD.}

\software{NumPy\footnote{\url{https://numpy.org}} \texttt{Ver 1.16.4} \citep{harris}, SciPy\footnote{\url{https://scipy.org}} \texttt{Ver 1.3.0} \citep{virtanen}, Matplotlib\footnote{\url{https://matplotlib.org}} \texttt{Ver 3.1.0} \citep{hunter} packages in Jupyter environment \citep{jupyter} for Python \texttt{3} \citep{rossum},
the SAO/NASA Astrophysics Data System\footnote{\url{https://ui.adsabs.harvard.edu/}}, 
e-Print arXiv\footnote{\url{https://arxiv.org/}}.}

\appendix

\section{\texttt{Surface Coverage} module} \label{simu}
\begin{figure}
    \centering 
    \includegraphics[width=.44\textwidth, clip, trim = 520 200 560 320]{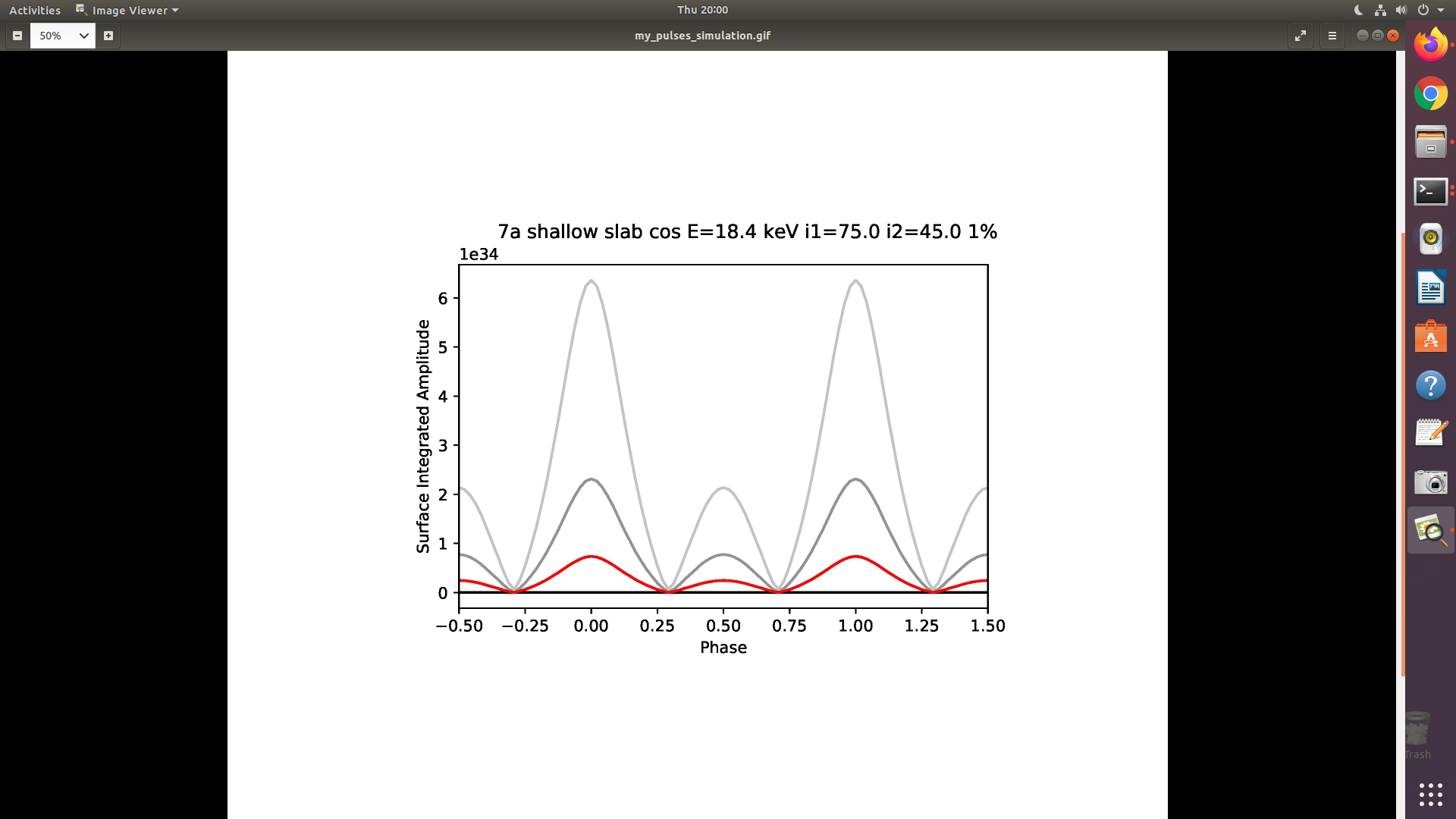}
    \includegraphics[width=.44\textwidth, clip, trim = 520 200 560 320]{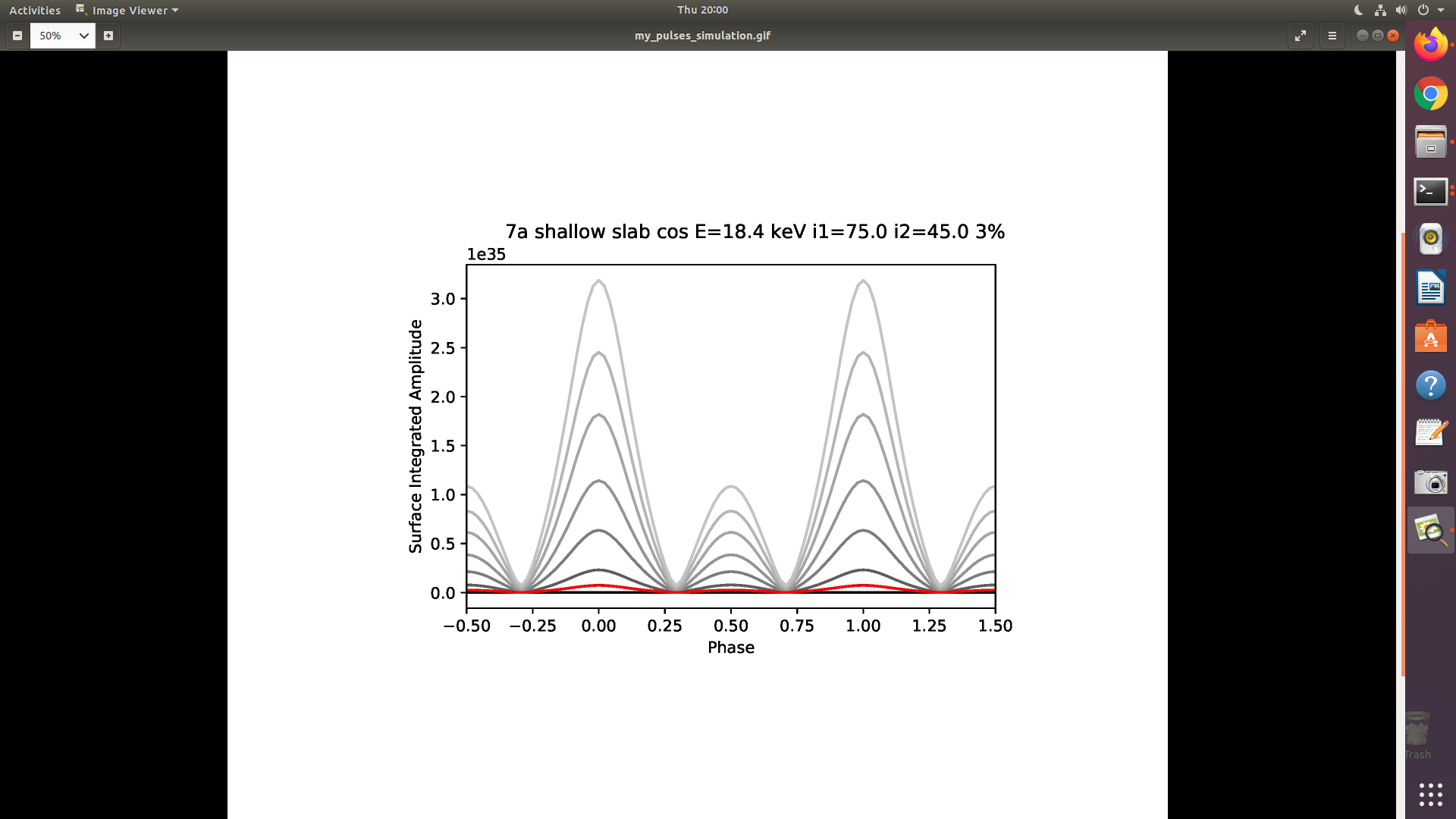}
    \includegraphics[width=.44\textwidth, clip, trim = 520 200 560 320]{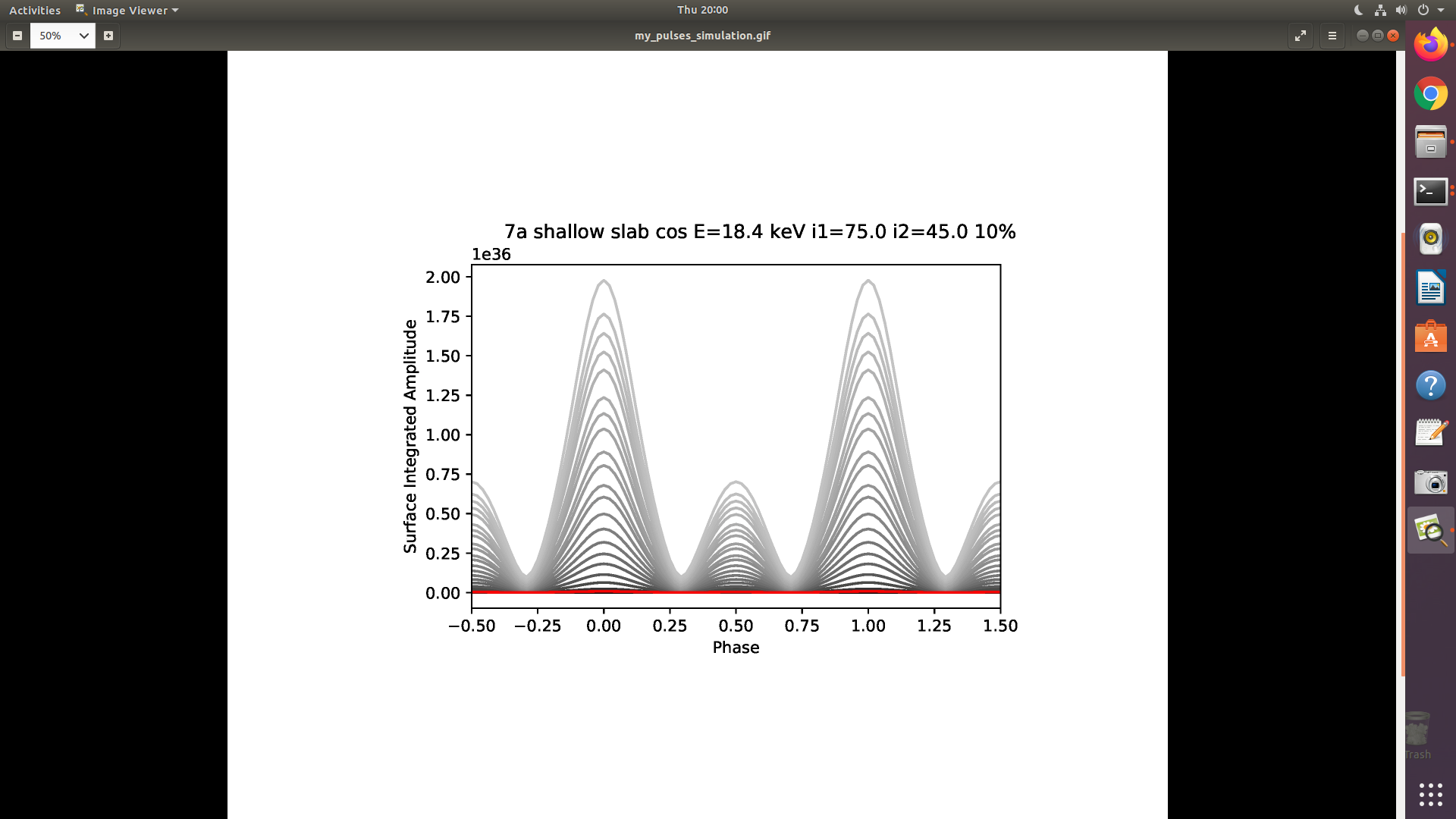}
    \includegraphics[width=.44\textwidth, clip, trim = 520 200 560 320]{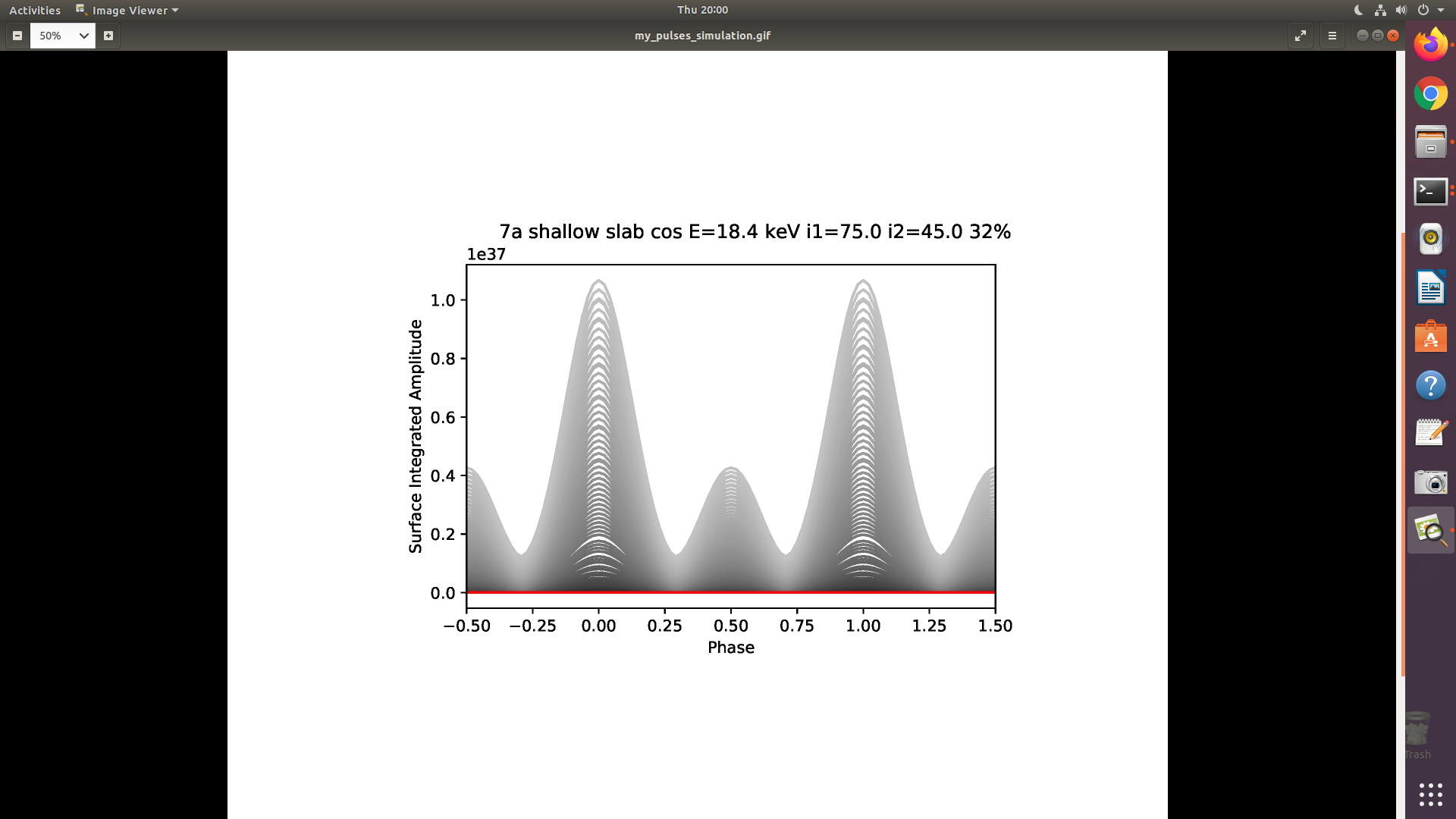}\\
    \includegraphics[width=.475\textwidth,trim= 0 238 970 258, clip]{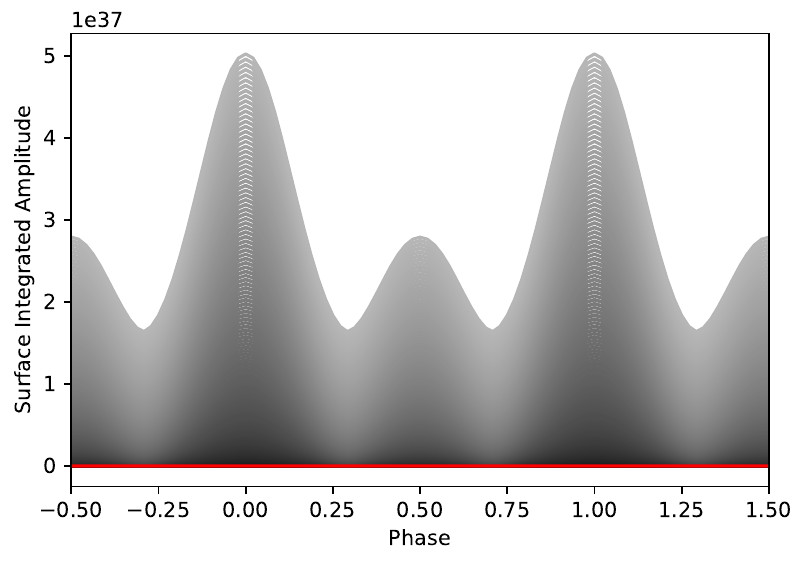}
    \includegraphics[width=.475\textwidth, trim= 0 0 0 20, clip]{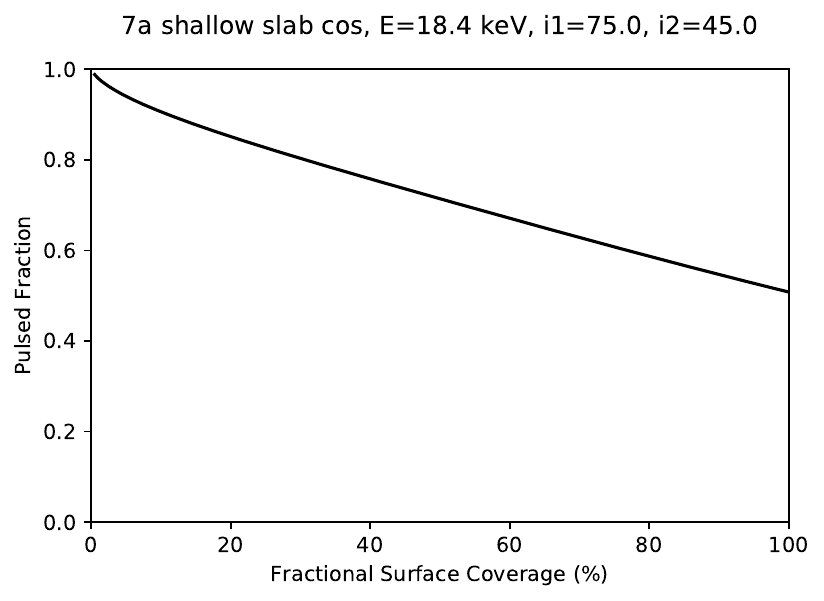}
    \caption{The \lq Phantom Figure' for beamed slab pulse profiles varying with fractional surface coverage of a flat hotspot. Panels provide simulated surface-integrated pulse profiles from 0\% up to (\textit{top left}) $1\%$, (\textit{top right}) $3\%$, (\textit{middle left}) $10\%$, (\textit{middle right}) $32\%$ and (\textit{bottom left}) 100\% surface coverage. The profiles at $100\%$ are reported at a coverage resolution of 0.5\% for pulsar inclinations $(i_1=75\textsuperscript{o}, i_2=45\textsuperscript{o})$ without light bending. The time slices depicted in this figure were obtained for logarithmically-spaced (with base 10) surface coverage fractions. The corresponding order of magnitude of the modeled flux shown in scientific notation above the $Y$-axis in each panel increases by one at each of these time instances, with the overall range remaining consistent with the typical, known span of X-ray pulsar luminosities. The black line denotes the extreme case of 0\% coverage and the red line corresponds to the standard emission area of 1 km$^2$. Grey lines depict increasing surface coverage for fainter shades. The curves do not flatten even for 100\% surface coverage due to beamed injection in the presence of a magnetic field. (\textit{bottom left}) The \lq Phantom Figure' (See Sec. \ref{sec:phantom}) with (\textit{bottom right}) its corresponding variation in the pulsed fraction with an increasing fractional surface coverage.}
    \label{fig:run}
    \label{phantom}
\end{figure}

\begin{figure}
    \centering 
    \includegraphics[trim=0 238 970 258, clip=true, width=.32\textwidth]{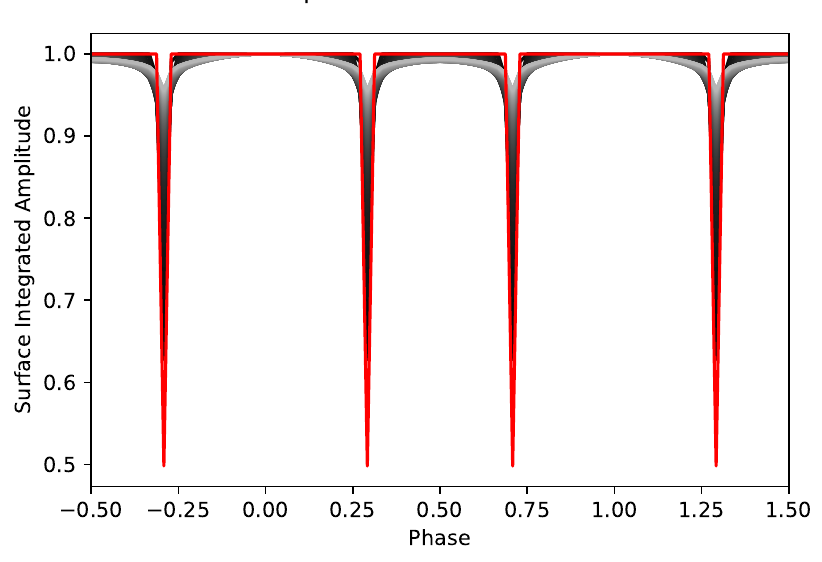} \hfill
    \includegraphics[trim=0 238 970 258, clip=true, width=.32\textwidth]{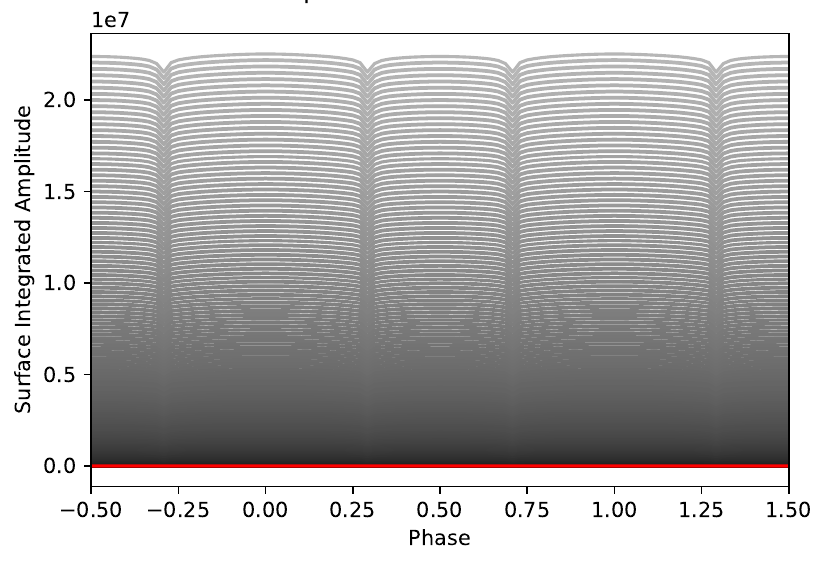} \hfill
    \includegraphics[trim=0 0 0 20, clip=true, width=.32\textwidth]{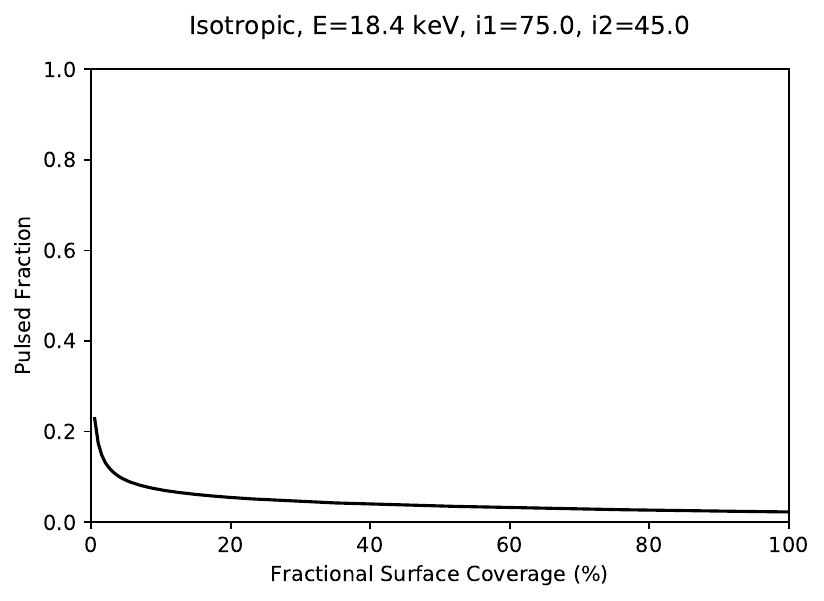}  \\
    \includegraphics[trim=0 238 970 258, clip=true, width=.5\textwidth]{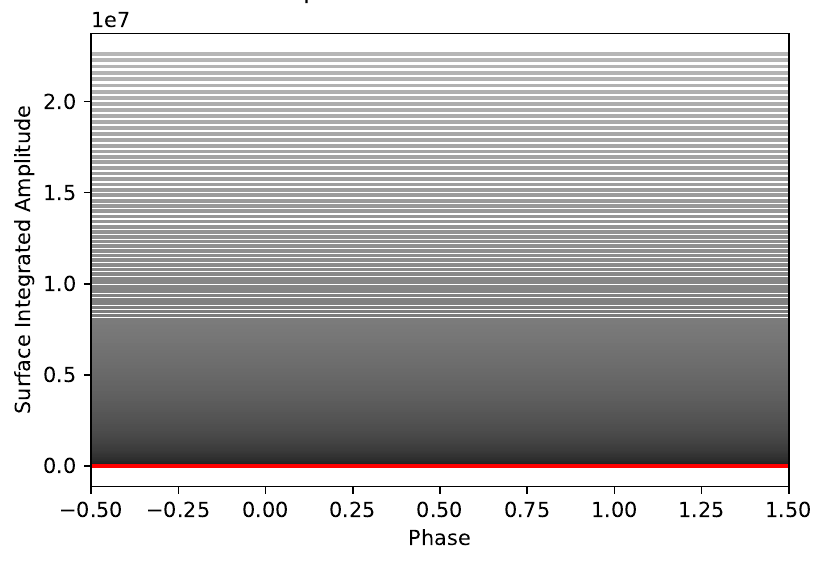} 
    \\
    \includegraphics[trim=0 238 970 258, clip=true, width=.32\textwidth]{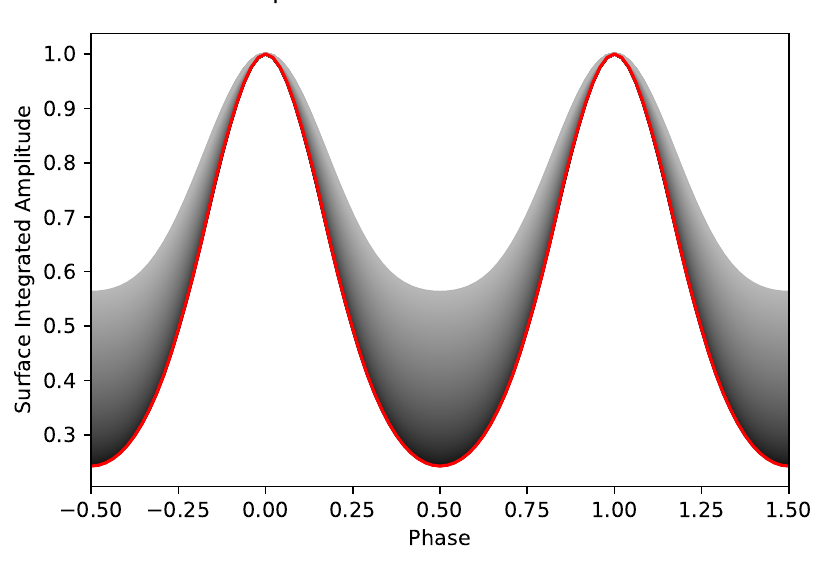} \hfill
    \includegraphics[trim=0 238 970 258, clip=true, width=.32\textwidth]{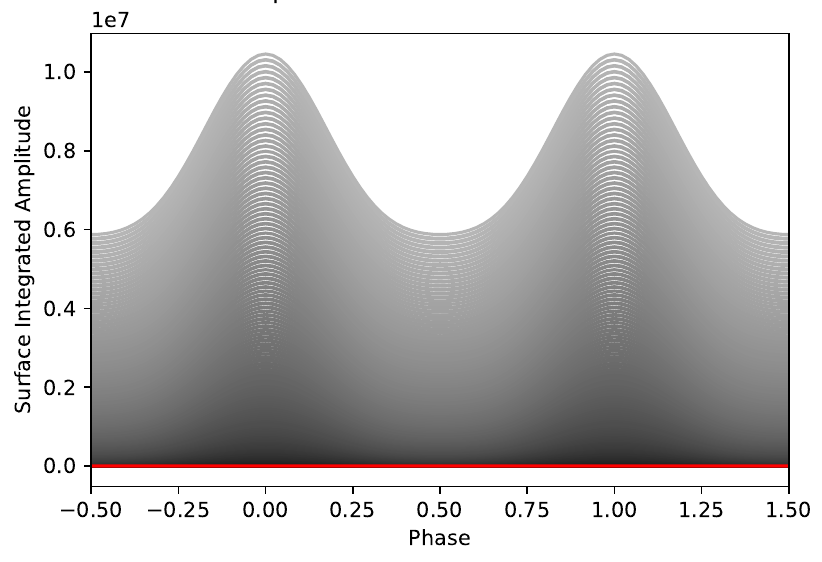} \hfill
    \includegraphics[trim=0 0 0 20, clip=true,width=.32\textwidth]{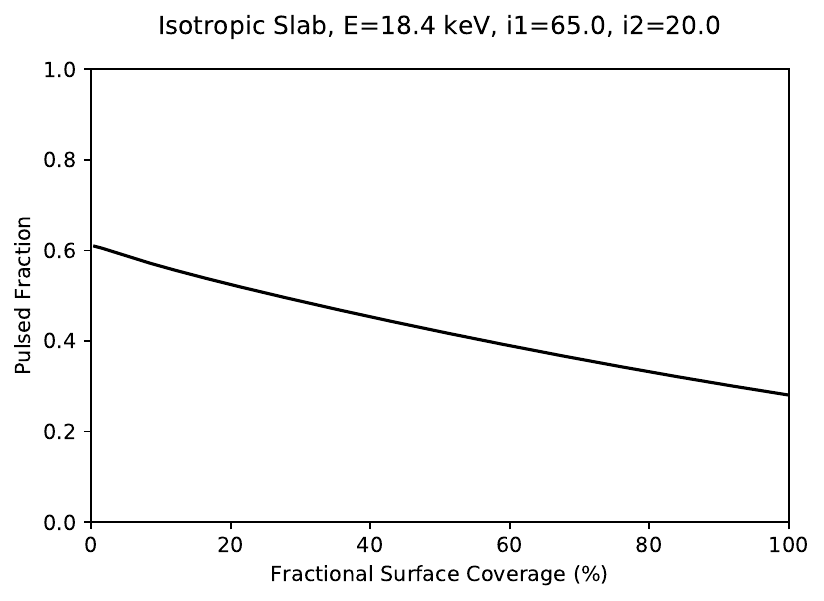} \\
     \includegraphics[trim=0 238 970 258, clip=true, width=.32\textwidth]{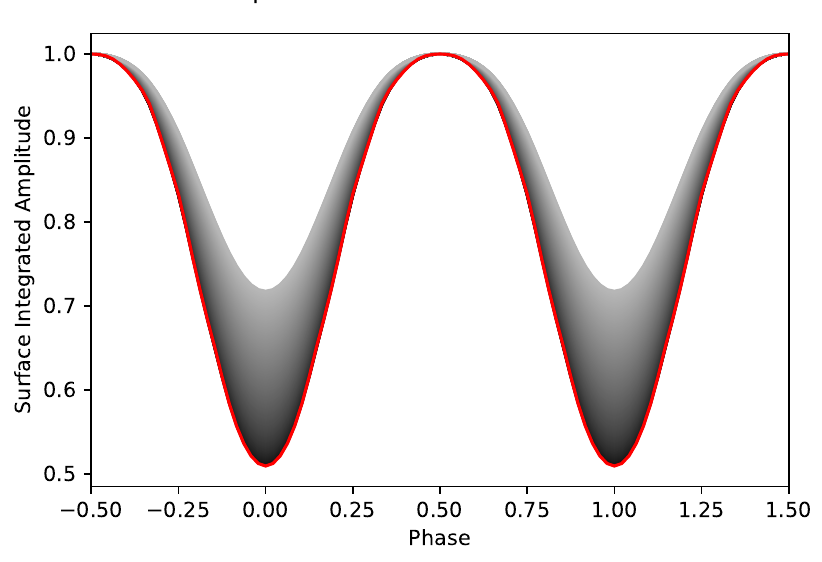} \hfill
    \includegraphics[trim=0 238 970 258, clip=true, width=.32\textwidth]{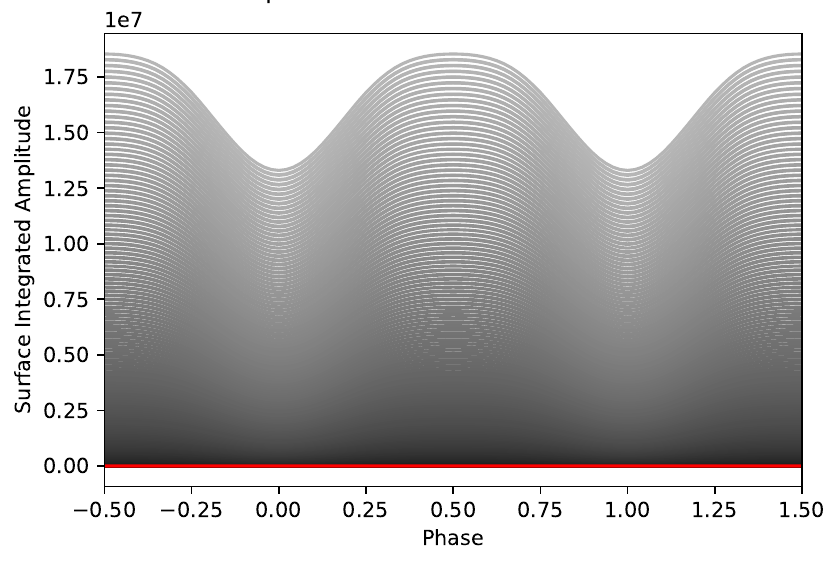} \hfill
    \includegraphics[trim=0 0 0 20, clip=true,width=.32\textwidth]{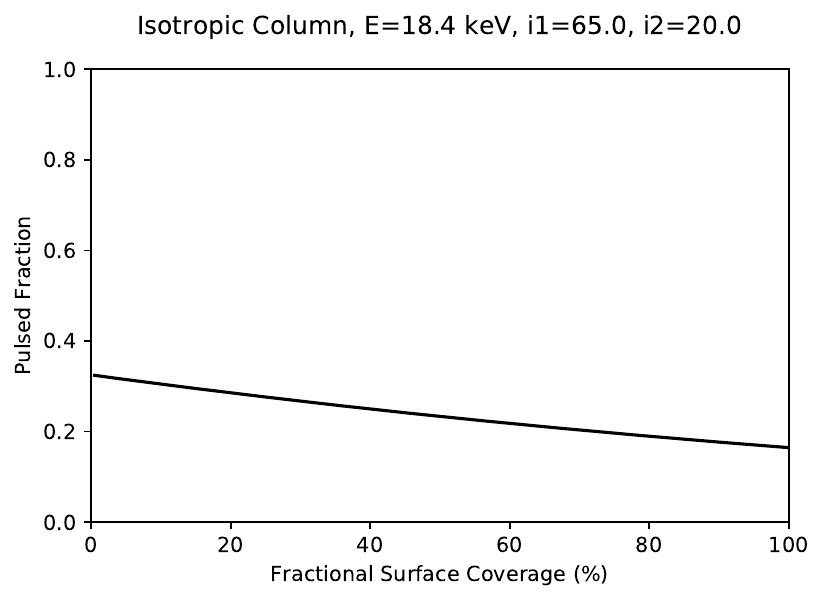}  \\
    \caption{Simulations for isotropic slab pulse profiles varying with a fractional surface coverage of a flat hotspot and the corresponding variation in the pulsed fraction with an increasing fractional surface coverage for different emission geometric projections. The top panel presents the \textit{(top left)} \lq The Dracula Fangs Diagram' i.e. normalized and \textit{(top middle)} non-normalized surface-integrated pulse profiles are isochors with 0.5\% change in fractional area coverage for a uniform emitter and pulsar inclination of ($75\textsuperscript{o},~45\textsuperscript{o}$) without light bending and \textit{(top right)} the corresponding variation in pulsed fraction with percentage surface coverage over the full range of $0-100\%$. Use of colours is the same as Fig. \ref{phantom}. The flux range in the middle column is of the order of 10$^7$ CGS units. \textit{(Second row)} Surface-integrated pulse profiles with varying percentage surface coverage for a perfectly uniform, isotropic emitter with the Centaurus X-3 ($i_1=65\textsuperscript{o}$, $i_2=20\textsuperscript{o}$) inclination. For these flat profiles, the pulsed fraction stays zero throughout. The third row presents the corresponding plots for a limb-darkened isotropic slab (with a cosine factor) and Centaurus X-3 pulsar inclination ($65\textsuperscript{o},~20\textsuperscript{o}$). The bottom row shows the respective plots for a limb-darkened isotropic column (with a sine factor) at the same inclination.}
    \label{fig:special_plots_isotropic}
    \label{dracula_fangs}
\end{figure}

Based on the prescription presented in Sec. \ref{spherical}, a new numerical \lq \texttt{Surface Coverage}' module is developed to provide results for a varying fractional coverage of the neutron star surface area by spherical caps. As an inherited feature from the \texttt{AXP4} code, the module can handle any pulsar inclination for 8 discrete values of X-ray photon energies and 4 emission geometries. A sample simulation run in progress is shown in Fig. \ref{fig:run}. The final result over the full range from 0\% -- 100\% at a high coverage resolution of 0.5\% is seen in Fig. \ref{phantom}. To test and verify the proper operation of the code, results are also obtained for other emission geometries like (i) uniform, isotropic emitter, (ii) isotropic slab, with an added cosine limb darkening \lq beaming' factor but without \citeauthor{MeszarosNagel1985a}'s
magnetic beaming for the other cases and (iii) isotropic column with a sine limb-darkening factor. These are shown in Fig. \ref{fig:special_plots_isotropic}. The $X$-axis of the last column in Fig. \ref{fig:special_plots_isotropic} depicts the percentage surface coverage made to vary the full range of $0\%-100\%$. In general, increasing the surface coverage increases the persistent, non-pulsed emission component and decreases the pulsed fraction.

\subsection{Special plots for a full 0 -- 100\% span of surface coverage} \label{sec:phantom}
The variation over the surface coverage fraction from 0\% to 100\% presents a series of similar-looking plots for various X-ray photon energies and emission geometries and inclinations provided as input. These have been named in this work to distinguish each series clearly from one another based on its overall resulting appearance to avoid ambiguity.

\subsubsection{The \lq Phantom Figures' for a beamed emitter} 
Fig. \ref{phantom} shows the pulse profiles for the full 0 -- 100\% range of compact stellar surface fractional coverage parsed at a 0.5\% coverage resolution. The black line indicates the extreme limiting case of 0\% coverage and the red line shows a standard hotspot of 1 km$^2$ area. Successively larger coverage spans are depicted by fainter grey lines. As expected, the pulsed fraction keeps decreasing with an increase in coverage as confirmed in the last column of Fig. \ref{fig:special_plots_isotropic}. Because the input emission function is beamed in the presence of a magnetic field, even the last profile corresponding to 100\% coverage never flattens to a limiting non-pulsed nature (See Sec. \ref{np}). Owing to their appearance, these have been christened the \lq Phantom Figures'\footnote{Similar to the \lq Butterfly Diagram' in the field of Solar Physics}.

\subsubsection{The \lq Dracula Fangs Diagram' for a uniform emitter} 

The diagrams in this paper can be compared with those produced for isotropic emitters for reference base. If the magnetic field and surface projection are removed, then the isotropic emission remains flat for all values of surface coverage as expected. Fig. \ref{fig:special_plots_isotropic} explores different cases of isotropic emitters namely, (i) uniform which provides a series of \lq Dracula Fangs Diagrams', (ii) with an additional cosine factor of projected slab emission for limb darkening and (iii) column emission with an additional sine factor for orthogonal emission for an orientation (75\textsuperscript{o}, 45\textsuperscript{o}) from \citeauthor{MeszarosNagel1985b}'s range followed by the (65\textsuperscript{o}, 20\textsuperscript{o}) orientation for the Centaurus X-3 pulsar.

\subsubsection{Non-pulsed flux component} \label{np}

\begin{figure}
    \centering
    \includegraphics[width=.48\textwidth]{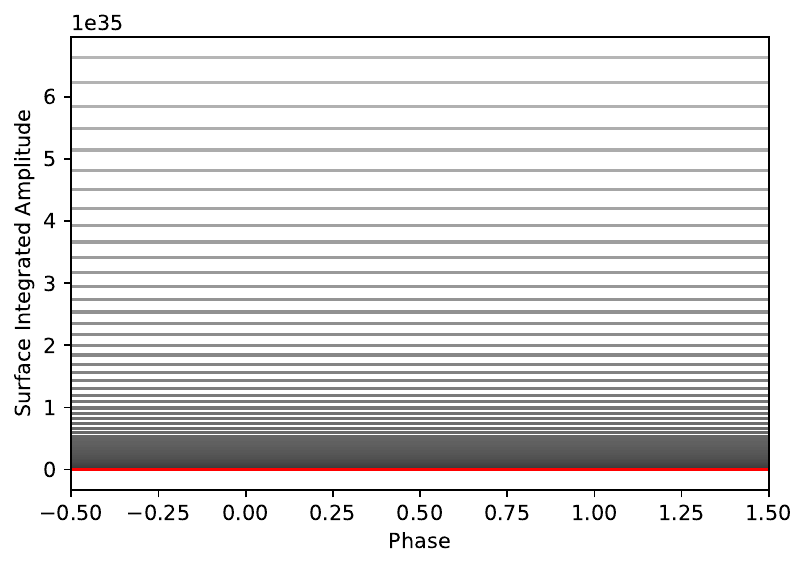}
    \includegraphics[trim=0 238 970 258, clip=true, width=.48\textwidth]{Ghost_Surface_Integrated_Pulse_Profiles_Vary_Coverage_Fraction_SphericalCorr_Isotropic_18.4_keV_65.0_20.0.pdf}
    \caption{(\textit{left}) Non-pulsed profiles with varying percentage surface coverage for the special case of $(i_1=90\textsuperscript{o},~i_2 = 0\textsuperscript{o})$ inclination without light bending at a coverage resolution of $0.5\%$. The profiles are the same for $(i_1=0\textsuperscript{o},~i_2 = 90\textsuperscript{o})$. The flux value is 10$^{35}$ ergs/s. (\textit{right}) Pulse profiles with varying percentage surface coverage for a uniform, isotropic emitter with the Cen X-3 ($i_1=65\textsuperscript{o}$, $i_2=20\textsuperscript{o}$) inclination. The color scheme in both the panels is the same as Fig. \ref{phantom}.}
    \label{fig:vary_M2}
    \label{fig:flat}
\end{figure}

\begin{figure}
    \centering
    \includegraphics[trim=0 0 0 20, clip=true,width=.8\textwidth]{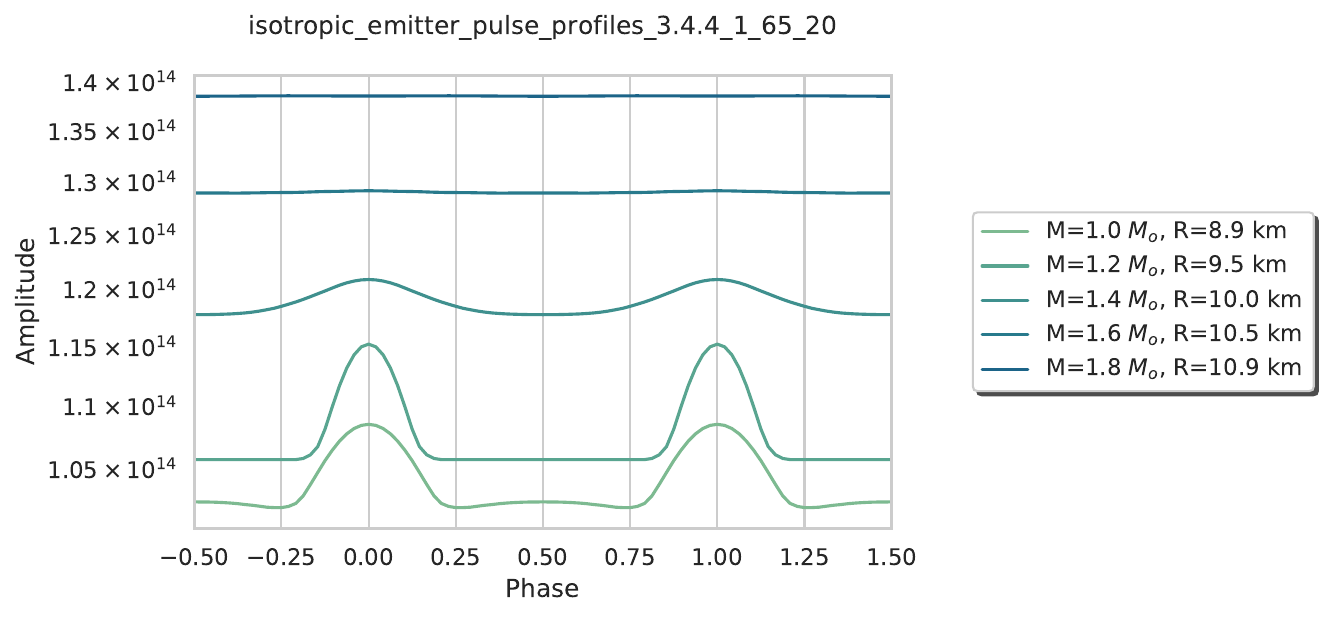}
    \caption{Pulse profiles for limb-darkened isotropic slabs with varying mass (and radius) of the neutron star with an inclination of ($i_1=65\textsuperscript{o}$, $i_2=20\textsuperscript{o}$).  Darker curves represent heavier stellar masses staggered by 0.2 M$_{\odot}$ starting from 1 M$_{\odot}$ for the lightest one. The radius is estimated assuming constant model density as calculated for the model $M=1.4$~M$_{\odot}, R = 10$ km case. The curves tend to non-pulsed behaviour approaching an Einstein ring for limiting mass values.}
    \label{fig:vary_M}
\end{figure}

The luminosity is primarily powered by the accretion process. As the nuclear burning lies in a stable regime, its contribution to the luminosity is insignificant (See \cite{2022arXiv220414185M} for a review). The emission coming from the other parts of the accretion flow \citep{bachhar2022timing} or by scattering of the X-rays from the central source by the stellar wind \citep{10.1111/j.1365-2966.2008.13251.x} can contribute to the un-pulsed continuum component.
In Fig. \ref{fig:flat}, we see that for an orthogonal rotator with the special case of source inclinations $(i_1=90\textsuperscript{o},  i_2 = 0\textsuperscript{o})$, the pulse profiles remain flat throughout the pulsar rotation, i.e. no pulsations are seen since the beams do not \textit{sweep} across the observer's line-of-sight. Further, as expected, it is also seen that the luminosity increases with surface coverage. This behaviour is confirmed by a corresponding flat pulsed fraction curve. This characteristic is exhibited for the $(i_1=0\textsuperscript{o},~i_2 = 90\textsuperscript{o})$ case, as well. As discussed in depth in the accompanying paper, Fig. \ref{fig:vary_M} shows the limiting case of the formation of an Einstein ring with increasing compactness of the neutron star for the Cen X-3 inclination.

\bibliography{sample631.bib}
\bibliographystyle{aasjournal}

\end{document}